\documentclass[fleqn,usenatbib,useAMS]{mnras}
\pdfoutput=1
\usepackage{graphicx}
\usepackage{amsmath, amssymb}
\usepackage[usenames,dvipsnames]{xcolor}
\usepackage[utf8]{inputenc}
\usepackage{ae,aecompl}
\usepackage[british]{babel}
\usepackage{mathptmx} 
\usepackage{balance}  
\usepackage{lastpage} 
\usepackage{fancyvrb} 

\DeclareMathAlphabet{\mathcal}{OMS}{cmsy}{m}{n}
\DeclareSymbolFont{greekletters}{OML}{cmm} {m}{it}
\DeclareMathSymbol{\epsilon}{\mathalpha}{greekletters}{"0F}
\DeclareMathSymbol{\theta}{\mathalpha}{greekletters}{"12}
\DeclareMathSymbol{\rho}{\mathalpha}{greekletters}{"1A}
\DeclareMathSymbol{\phi}{\mathalpha}{greekletters}{"1E}
\DeclareMathSymbol{\psi}{\mathalpha}{greekletters}{"20}
\DeclareMathSymbol{\varepsilon}{\mathalpha}{greekletters}{"22}
\DeclareMathSymbol{\vartheta}{\mathalpha}{greekletters}{"23}
\DeclareMathSymbol{\varrho}{\mathalpha}{greekletters}{"25}

\defcitealias{tejeda17}{TGRM+17}
\defcitealias{guillochon13}{GRR13}

\newcommand{\en}{{\mathcal{E}}} 
\newcommand{\lz}{{\ell_z}} 
\newcommand{\q}{{\mathcal{Q}}} 
\newcommand{\aBH}{{a^\star}} 
\newcommand{\tup}{t_{10\rightarrow 100}} 
\newcommand{\tdown}{t_{100\rightarrow 10}} 

\title[Tidal disruptions by rotating black holes]{Tidal disruptions by rotating black holes: effects of spin and impact parameter}

\author[E. Gafton \& S. Rosswog]{Emanuel Gafton$^{1,2}$\thanks{E-mail: \href{mailto:ega@ing.iac.es}{ega@ing.iac.es} (EG), 
\href{mailto:stephan.rosswog@astro.su.se}{stephan.rosswog@astro.su.se} (SR).} and   
Stephan Rosswog$^{1}$
 \\
$^{1}$Department of Astronomy and Oskar Klein Centre, Stockholm University, AlbaNova, SE-10691 Stockholm, Sweden\\
$^{2}$Isaac Newton Group of Telescopes, Calle {\'A}lvarez de Abreu 70, ES-38700 Santa Cruz de La Palma, Spain
}

\date{Accepted 2019 May 29. Received 2019 May 23; in original form 2019 March 21}

\pubyear{2019}

\begin{document}
\label{firstpage}
\pagerange{\pageref{firstpage}--\pageref{LastPage}}
\maketitle

\begin{abstract}%
We present the results of relativistic smoothed particle hydrodynamics simulations 
of tidal disruptions of stars by rotating supermassive black holes, for a wide range 
of impact parameters and black hole spins. 
For deep encounters, we find that: relativistic precession creates debris geometries 
impossible to obtain with the Newtonian equations; part of the fluid can be launched 
on plunging orbits, reducing the fallback rate and the mass of the resulting accretion 
disc; multiple squeezings and bounces at periapsis may generate distinctive X-ray 
signatures resulting from the associated shock breakout; disruptions can occur inside 
the marginally bound radius, if the angular momentum spread launches part of the 
debris on non-plunging orbits. 
Perhaps surprisingly, we also find relativistic effects important in partial disruptions, 
where the balance between self-gravity and tidal forces is so precarious that otherwise 
minor relativistic effects can have decisive consequences on the stellar fate. 
In between, where the star is fully disrupted but relativistic effects are mild, the 
difference resides in a gentler rise of the fallback rate, a later and smaller peak, 
and longer return times. However, relativistic precession always causes thicker debris 
streams, both in the bound part (speeding up circularization) and in the unbound part 
(accelerating and enhancing the production of separate transients). We discuss various 
properties of the disruption (compression at periapsis, shape and spread of the energy 
distribution) and potential observables (peak fallback rate, times of rise and decay, 
duration of super-Eddington fallback) as a function of the impact parameter and the 
black hole spin.
\end{abstract}


\section{Introduction}
As observational evidence for the tidal disruption of stars by supermassive
black holes (SMBH) is mounting, the question remains how such observations
can help determine black hole (BH) properties such as mass and spin (\citealp*{mockler19}). 
While analytical calculations may be able to provide a strikingly accurate 
description of some features of such an event, including the famous $t^{-5/3}$ 
decay of the fallback rate (\citealp{rees88,phinney89}), a relativistic tidal disruption event (TDE) is
a highly non-linear outcome of the interplay between the stellar hydrodynamics
and self-gravity, tidal accelerations from the black hole, radiation, potentially magnetic
fields and -- in extreme cases -- nuclear reactions. To date, systematic  numerical
studies of relativistic TDEs are still lacking.

In this paper, we shall use the formalism introduced by \citeauthor{tejeda17} 
(\citeyear{tejeda17}, henceforth \citetalias{tejeda17}), ``relativistic hydrodynamics 
with Newtonian codes'', to perform an extensive  study of canonical tidal disruption 
events around spinning black holes. The method combines an exact relativistic 
description of the hydrodynamical evolution of a fluid in Kerr spacetime
with a quasi-Newtonian treatment of the fluid’s self-gravity. The only similar 
study of the parameter space we are aware of belongs to
\citeauthor{guillochon13} (\citeyear{guillochon13}, henceforth 
\citetalias{guillochon13}), who analysed the dependency of Newtonian TDEs on the 
penetration factor $\beta$, for two types of polytropes (with polytropic exponent 
$\gamma=4/3$ and $5/3$), and with $\beta$ ranging from $0.6$ to $4$ (for $\gamma=4/3$) 
and from $0.5$ to $2.5$ (for $\gamma=5/3$).

High-resolution relativistic simulations of TDEs have only become feasible in recent years. 
Since the seminal paper by \citet{laguna93}, who tackled for the first time disruptions of 
Main-Sequence (MS) stars around Schwarzschild BHs, relatively few studies have 
continued the numerical investigation of relativistic TDEs: \citet{kobayashi04} 
repeated the simulations of \citet{laguna93}  ($\beta=1$, $5$ and $10$) using 
essentially the same numerical method (\citealp*{laguna93a}), and additionally 
treated the disruption of a helium star with $\beta=1$, with the goal of predicting 
the X-ray and gravitational wave signatures from such TDEs,
while \citet{bogdanovic04} used the same code to study a canonical disruption with $\beta=1.2$ 
and calculate the H~$\alpha$ emission-line luminosity of the resulting disc and debris tail.
\citet{cheng14b} presented three disruptions of MS stars with $\beta=1$ 
around SMBHs of masses $M=10^5$, $10^6$ and $10^7~{\rm M}_{\sun}$,  
and of two of white dwarfs (WD) with $\beta=5$, $6$ around a $10^5~{\rm M}_{\sun}$ 
BH, focusing on the influence of relativistic effects on the fallback rate of the bound debris.
Other authors have chosen to focus exclusively on the effect of General Relativity (GR) 
on WD disruptions by intermediate-mass black holes (IMBH), since they pose less 
of a computational challenge \citep{haas12,cheng14,shiokawa15}.

In recent years, another class of simulations, combining particle codes for the 
disruption part of a TDE and fixed metric general-relativistic Eulerian codes for 
the disc formation part, have improved our understanding of the process of 
accretion disc formation in TDEs (e.g., \citealp{sadowski16b}). However, these 
studies have generally focused on a very reduced set of initial parameters (one 
or two encounters being the norm), and have invariably approached TDEs on 
elliptical orbits in order to alleviate the tremendous scale problem of a typical 
parabolic encounter. These studies are complemented by pseudo-relativistic 
particle simulations of the disruption of stars on elliptical orbits, followed by 
the subsequent circularization (e.g., \citealp{bonnerot15}; \citealp*{hayasaki13,hayasaki15}).

The literature of TDEs in Kerr spacetime is substantially sparser: aside from some 
older analytical and semi-analytical studies \citep{luminet85,frolov94,kesden12}, 
only two numerical studies have included the effects of the BH spin:
a) \citet{haas12} presented six ultra-close TDEs of a $1~{\rm M}_{\sun}$ WD by a
$10^3~{\rm M}_{\sun}$ Kerr IMBH with spin parameters $\aBH=0$ and $\aBH=0.6$;
b) \citet*{evans15} presented nine ultra-close TDEs of a solar-type star by a $10^5~{\rm M}_{\sun}$ 
Kerr IMBH with spin parameters $\aBH=0$ and $\pm 0.65$.
In both cases, the parameter range has been fine-tuned with the periapsis being close to the 
marginally bound radius of the BH, i.e., at the very limit of where a TDE could possibly take place.

All these studies necessarily focused on restrictive subsets of the parameter space 
(often chosen to reduce the otherwise prohibitive computational burden), and while 
they have answered important specific questions about the impact of GR on the 
disruption process and in particular on the circularization, they cannot provide 
an overview of Kerr TDEs across the range of impact parameters and spins. To date, 
this has only been achieved by (semi-)analytical means, by \citet{kesden12}.

To our knowledge, this paper is the first systematic numerical study of TDEs with 
relativistic hydrodynamics around Kerr BHs. Our goal is to analyse the disruption 
process from the initial approach until the second periapsis passage, and to: 
a) compare our findings with previous results (both for Newtonian and  Kerr BHs) 
and with standard expectations based on theoretical arguments;
b) determine the effects of GR in general, and of the BH spin in particular, 
on the various stages of the disruption;
c) present a unified picture of tidal disruptions in Kerr spacetime 
(since we also treat the particular case $\aBH=0$, we implicitly include 
Schwarzschild BHs in our analysis).

The importance of general relativistic effects has recently been reviewed by
  \citet{stone19}, whereas results from simulations are summarized by \citet{lodato15}
  and recent observational advances on TDEs can be found in \citet{komossa15}.

\section{Method}
\subsection{Code and equations}
All simulations presented in this paper used a modified version of a 
Newtonian Smoothed Particle Hydrodynamics (SPH) code (\citealp{rosswog08b}).
For general reviews of the method see \citet{monaghan05}; 
\citeauthor{rosswog09b} (\citeyear{rosswog09b}, \citeyear{rosswog15c}).
We modified the Newtonian accelerations due to the BH, the fluid self-gravity and 
the pressure forces according to the ``Generalized Newtonian'' prescription 
introduced in \citetalias{tejeda17}. This approach combines exact hydrodynamic 
and black hole forces in Kerr spacetime with a \mbox{(quasi-)}Newtonian 
treatment of the stellar self-gravity that reduces to the Newtonian formulation 
far from the BH, and that ensures hydrostatic equilibrium of  the pressure 
and self-gravity forces in the rest frame of the star.
This method was carefully scrutinized in \citetalias{tejeda17} and showed
excellent agreement with the few existing relativistic TDE simulations and consistency with
fundamental principles such as coordinate invariance.

\subsection{Parameter space}
\begin{table}
\centering
\caption{Overview of the SPH simulations presented in this paper. The parameter 
space spans the $26\times 5$ possible combinations of $\beta$ and $\aBH$, 
plus 26 additional control simulations with a Newtonian BH.}
\label{table:overview}
\begin{center}
 \begin{tabular}{lll}
 \hline
  Quantity & Symbol & Value(s) \\
 \hline
 BH mass & $M$  & $10^6~{\rm M}_{\sun}$\\
 Stellar mass & $m_\star$ & ${\rm M}_{\sun}$\\
 Stellar radius & $r_\star$ & ${\rm R}_{\sun}$\\
 Tidal radius & $r_{\rm tid}$ & $r_\star \left(M/m_\star\right)^{1/3}$\\
 Initial separation & $r_0$ & $5~r_{\rm tid}$ \\
 Polytropic index & $\gamma$ & $5/3$ \\
 Adiabatic exponent & $\gamma_{\rm ad}$ & $5/3$ \\
 Impact parameter & $\beta$ & \{0.50, 0.55, 0.60, 0.65, 0.70, \\
 && \phantom{\{}0.75, 0.80, 0.85, 0.90, 0.95, \\
 && \phantom{\{}1.0, 1.1, 1.2, 1.3, 1.4, 1.5, \\
 && \phantom{\{}2, 3, 4, 5, 6, 7, 8, 9, 10, 11\}\\
 BH spin & $\aBH$ & $\{-0.99$, $-0.5$, $0$, $0.5$, $0.99\}$ \\
 SPH particles & $N_{\rm part}$ & 200\,642 \\
 \hline
 \end{tabular}
\end{center}
\end{table}

We perform a large set of simulations to systematically explore the 
effects of the BH spin on the morphology and properties of the debris 
resulting from tidal disruptions, a study that, to our knowledge, has 
not been performed before due to the lack of suitable numerical tools.

Throughout this paper, we will refer to the typical quantities involved in a
tidal disruption with the following abbreviations:
the black hole mass $M$; the stellar mass $m_\star$ and radius $r_\star$;
the periapsis $r_{\rm p}$;
the tidal radius $r_{\rm tid}\equiv r_\star(M/m_\star)^{1/3}$;
the impact parameter or penetration factor $\beta\equiv r_{\rm tid}/r_{\rm p}$;
the gravitational radius of the BH $r_{\rm g}\equiv GM/c^2$;
the BH angular momentum $J$, represented through the dimensionless spin 
parameter $\aBH\equiv Jc/GM^2$ ranging from $-1$ to $1$, with the convention 
that $\aBH > 0$ for prograde orbits and $\aBH < 0$ for retrograde orbits.

We focus our study to canonical tidal disruptions of solar-type stars
($m_\star={\rm M}_{\sun}$, $r_\star={\rm R}_{\sun}$) on parabolic orbits (eccentricity $e=1$, 
specific mechanical energy $\en=0$) that are disrupted by 
$M=10^6~{\rm M}_{\sun}$ supermassive black holes.
We perform simulations for various impact 
parameters, ranging from $\beta=0.5$ (corresponding to a periapsis 
distance $r_{\rm p}/r_{\rm g}\approx 94$)
to $\beta=11$ (corresponding to a periapsis distance $r_{\rm p}/r_{\rm g}\approx 4.3$); 
note that the limit for disruption is the marginally bound orbit (\citealp*{bardeen72}), spanning
from $r_{\rm p}/r_{\rm g}=1.21$ (or $\beta\approx 38.9$) for
$\aBH=0.99$, to $r_{\rm p}/r_{\rm g}=4$ (or $\beta\approx 11.7$)
for $\aBH=0$, to $r_{\rm p}/r_{\rm g}\approx 5.8$ (or
$\beta\approx 8.1$) for $\aBH=-0.99$, beyond which the star is
expected to plunge directly into the BH. For each individual
$\beta$ we run five simulations around Kerr black holes with spin
parameters $\aBH=0$, $\pm 0.5$, $\pm 0.99$, plus one simulation
around a Newtonian black hole, bringing the total number of
simulations up to 156\footnote{A complete table with the
resulting snapshots, either a) at the end of the simulation, or b)
before the first SPH particle enters the event horizon, is available online, at
\href{http://compact-merger.astro.su.se/~ega/tde}{http://compact-merger.astro.su.se/{\textasciitilde}ega/tde}.}. 

Table \ref{table:overview} summarizes the parameter range of our simulations.

The choice of a single black hole mass and stellar type for all simulations was 
imposed by the otherwise unwieldy parameter space that TDEs can span. 
We chose to focus our analysis on $\beta$ and $\aBH$, since the former is 
the parameter that most affects the qualitative outcome of the TDE, while 
the latter is one of the most basic yet so far unexplored relativistic parameters.
The only other non-trivial dependence is on the stellar structure, which 
in the case of a polytropic model is related to the polytropic exponent $\gamma$. 
For this work we chose to focus only on $\gamma=5/3$, so as to have a 
manageable parameter space and to use the same equilibrium star as an 
initial condition in all the simulations.
We also set up the initial star as non-rotating, while keeping in mind that 
stellar spin may have a non-negligible influence of the fallback rates 
(\citealp*{golightly19}; \citealp*{kagaya19}).

All the other quantities that describe a TDE (and in particular potential observables 
such as the peak fallback rate and time to peak) are expected to obey a simple 
scaling relation with the black hole mass and with the stellar mass and radius 
(see \citetalias{guillochon13}); most of them, however, have a non-trivial 
dependence on $\beta$.
Also, tidal disruption event rates scale weakly with the black hole mass 
(typically as $\propto M^{-0.25}$, e.g., \citealp{wang04}), but exhibit 
a stronger dependence on $\beta$ (typically scaling as $\propto \beta^{-1}$ for 
$\beta>1$, e.g., \citealp{rees88}).

\subsection{Initial conditions}
The SPH particles ($200\,642$ for all the simulations) are initially distributed 
on a close-packed hexagonal lattice so as to fulfil the Lane--Emden equations 
for a $\gamma=5/3$ polytrope, and are then relaxed with damping into 
numerical equilibrium (see \citealp*{rosswog09} for details).  The relaxed star 
is then placed on a parabolic orbit around the BH, using the equations derived 
in Appendix A2 of \citetalias{tejeda17}. The initial separation from the BH is 
$r_0=5\/r_{\rm tid}$, in order for the relaxed, spherically-symmetric star to be 
a valid initial condition, as discussed in Section 1 of \citetalias{tejeda17}. 
For all the simulations, we use an off-equatorial trajectory with initial latitude 
$\theta_0=0.5\pi$ (i.e., starting on the equatorial plane) and minimum latitude 
$\theta_{a'}=0.1\pi$ (we follow the angle conventions from the Appendix A2 of 
\citetalias{tejeda17}).  The initial azimuthal angle is $\phi_0=0$, and the star 
is imparted a positive initial azimuthal velocity.
During the simulation, the gas is evolved with a $\gamma_{\rm ad}=5/3$ 
equation of state, which is appropriate for a gas-pressure dominated fluid 
(e.g., \citealp{chandrasekhar39}).

\subsection{Postprocessing}
Most of the simulations are stopped $\sim 60$ hours ($\simeq 2.5$~days) 
after the start of the simulation, or $\approx 57.5$ hours after the periapsis passage 
(1 hour is comparable to the dynamical time scale of the initial polytrope and 
to the periapsis crossing time). There are two exceptions:

a) In the case of the deepest encounters ($\beta \gtrsim 9$), part of the tidal debris 
is already launched on plunging orbits (as will be shown later on) during the 
disruption itself, which poses numerical challenges long before reaching the 
stopping time of the other simulations. In order to be consistent and avoid the 
second periapsis passage altogether, we stop these simulations just before the 
first SPH particle on a plunging orbit enters the event horizon (at the times given 
in the upper part of Table~\ref{table:stopped}). Since the star is already fully 
disrupted at this early time, and there is no surviving core to exert its gravitational 
influence over the tidal tails, the particles are already moving on essentially ballistic 
trajectories and therefore the evolution up to the second periapsis passage may be 
accurately predicted based on geodesic motion alone.

b) In the case of the partial disruptions, due to the computational expense of 
evolving the surviving core, we stop the simulations once the star is outside the 
tidal radius of the black hole, and the evolution of energy and angular momentum 
in most of the material in the tidal tails slows down (at the times given in the lower 
part of Table~\ref{table:stopped}).

\begin{table}
\centering
\caption{Time at which the simulations were stopped, when different from 60.28 hours.}
\label{table:stopped}
\begin{center}
 \begin{tabular}{lllllll}
 \hline
 \hline
 Gravity & Spin & $\beta$ / t [h] &&&\\
 \hline
 Kerr & $-0.99$ & 9 / 2.59 &  10 / 2.56 &  11 / 2.55&\\ 
 Kerr & $-0.5$ & 9 / 2.59 &  10 / 2.55 &  11 / 2.54& \\
 Kerr & $0$ &  &  10 / 2.54 &  11 / 2.53& \\
 Kerr & $0.5$ & &   &  11 / 2.54 &\\
 Kerr & $0.99$ &  &   &  11 / 58.08 &\\
 Newton &  & 9 / 55.82 &  10 / 32.89 &  11 / 26.67& \\
 \hline
 Kerr & $-0.99$ & 0.50 / 48.54 & 0.55 / 51.09 & 0.60 / 43.52&\\
 Kerr & $-0.5$ & 0.50 / 47.29 & 0.55 / 50.54 & 0.60 / 32.16 & \\
 Kerr & $0$ &   0.50 / 49.21 & 0.55 / 37.23 && \\
 Kerr & $0.5$ & 0.50 / 48.42 &  0.55 / 50.66 & 0.60 / 60.04 &\\
 Kerr & $0.99$ &  0.50 / 49.60 & 0.55 / 37.33 & 0.60 / 31.97\\
 Newton &  & 0.50 / 12.25 & 0.55 / 12.87 & 0.60 / 53.31 &\\
 \hline
 \end{tabular}
\end{center}
\end{table}

After the simulations are stopped, we perform the following post-processing 
operations on the resulting snapshots:

\begin{enumerate}
\item extract the positions, densities, internal energies, and calculate the 
constants of motion (specific mechanical energy $\en$, specific angular momentum 
$\lz$, Carter constant $\q$, as defined in Appendix A1 of \citetalias{tejeda17}), 
for all of the particles;
\item based on $\en$, $\lz$, $\q$, compute the turning points $r_{\rm p}$ and 
$r_{\rm a}$ as the largest two of the four real roots of Eq.~(A18) in \citetalias{tejeda17}; 
when, instead, two of the roots are complex conjugates, there are no turning points; 
since in all such cases the radial velocity is negative, the particles in question are 
on plunging orbits, and as such are not considered for the computation of the 
$\dot{M}$ curve; in all these cases (for relativistic simulations with $\beta\gtrsim 9$), 
the plunging time down to the event horizon is less than an hour from the first periapsis passage;
\item we integrate the radial equation of motion to periapsis (for inbound particles) 
or to apapsis and then back to periapsis (for outbound particles) in order to calculate 
the fallback time; the histogram of this quantity gives the $\dot{M}$ curve;
\item in the case of partial disruptions, we determine which particles belong to the 
self-bound mass, and which to the bound and unbound tidal tails, following the 
iterative procedure described in \citetalias{guillochon13} and used before in 
\citet{gafton15}. Since the self-bound mass is a re-collapsing stellar core escaping 
the SMBH at high velocities,  the $\dot{M}$ curve is computed only with the particles 
from the bound tail, not from all the particles with $\en<0$.
\end{enumerate}

In order to plot snapshots of the tidal debris and analyse the possible morphological 
classes, we apply the following transformation on the data. Since the orbits are not 
equatorial (i.e., the Carter constant is not zero and therefore the motion is not 
confined to the equatorial plane, although the star does start at $\theta_0=\pi/2$), 
the plane of the debris is not $z=0$, and so plotting $y(x)$ or other projections 
onto the usual Cartesian axes leads to the particle distribution appearing distorted. 
This is particularly relevant for the close encounters in Kerr, where nodal precession 
yields a non-planar particle distribution.
Our solution is to fit a plane to the entire particle distribution (through linear regression), 
and then project the particles onto that plane. This is the simplest solution analogous 
to plotting $y(x)$ in a typical Newtonian simulation with the star on the equatorial ($z=0$) plane.

In addition, since the position, size, distance from the black hole and orientation of 
the disrupted star in the final snapshot all vary greatly (without being relevant for 
the morphology itself), we shift the origin of the coordinate system to the centre of 
mass of the debris, and express the distances in relativistic units of $r_{\rm g}$. 
We also rotate all snapshots to have the bound tail in the lower-left corner, and the 
unbound tail in the upper-right corner, by finding the slope $m$ of the line passing  
through the centres of mass of the unbound and bound debris, and then rotating 
the particle distribution by $\pi/4-\arctan(m)$. We will refer to these transformed 
Cartesian-like coordinates as $\tilde{x}$ and $\tilde{y}$, and they will be used as 
the $x$ and $y$ axes of our plots.

\subsection{Comparing Newtonian and relativistic disruptions}
As discussed at length by \citet{servin17}, comparing simulations of
Newtonian and relativistic TDEs is not straightforward, because there
are various mappings that all reduce to the Newtonian limit far from
the black hole. 

The obvious choice, generally adopted throughout the literature, is
that of considering orbits with equal periapsis distances (and hence
equal $\beta$). This amounts to comparing what happens to a star on a
parabolic orbit when approaching the black hole at the same minimum
distance in the two gravity theories. We believe this motivation to be
reasonable, and more relevant than the one given in the mentioned
paper (where it is linked to the circumference of a circular orbit
with the same periapsis as the parabolic orbit, which we agree is not
``particularly useful''). Another mapping proposed by \citet{servin17}
is that of orbits experiencing equal tidal forces at periapsis; this
is likely to yield the most similar disruptions, and they present an
analytical relationship between the usual impact parameter, $\beta$,
and the adjusted one, $\beta_{\rm N}$. 

In our study we will analyse the dependence of many quantities on
$\beta$. We will directly apply the first mapping, which is the most
prevalent throughout the literature and simplifies the comparison of
our results with previous ones, but we note that all our results can
be translated to the third mapping by a relatively simple scaling from
$\beta$ to $\beta_{\rm N}$, as given by Eq.~(23) of
\citet{servin17}. 

\section{Stages of a tidal disruption}
\subsection{Approach}

As the star enters the tidal radius on its approach to periapsis, the tidal force 
becomes comparable to the pressure and self-gravity. \citet*{lodato09} proposed 
a simple analytical model, usually referred to as the ``frozen-in model'', which 
assumes that all gravitational and hydrodynamic interactions within the fluid 
cease at a fixed point (originally taken to be the periapsis; \citet*{stone13} and 
\citetalias{guillochon13} proposed using instead the point of exit from being 
within the tidal radius), so that the fluid continues to move on ballistic trajectories. 
In spite of its simplicity, the model has resulted in fairly accurate predictions for 
the energy distribution and the mass return rates.

Since the frozen-in model posits the lack of hydrodynamic and self-gravitational 
interactions within the fluid, there is no mechanism for the exchange of energy 
and angular momentum, and therefore the constants of motion are ``frozen-in'' 
to the values they had at the fixed point. In particular, the specific orbital energies 
$\en$ are given by the  gradient of the BH potential ($\Phi_{\rm N}=-GM/r$ for 
the Newtonian case) at the fixed point, $r=r_{\rm fix}$; this can be estimated 
through a Taylor expansion across the star around $r_{\rm fix}$ as
\begin{equation}\label{eq:espread}
\Delta \en\simeq k_\en \beta^n \frac{GMr_\star}{{r_{\rm tid}}^2},
\end{equation}
with $k_\en$ being a constant of order unity related to the stellar structure and 
rotation. If $r_{\rm fix}=r_{\rm p}$ (\citealp{lodato09}), then $n=2$; if, however, 
$r_{\rm fix}=r_{\rm tid}$ (\citealp{stone13}; \citetalias{guillochon13}), then 
$n=0$ and the energy spread is independent of $\beta$, being solely determined 
by the stellar structure and the BH mass. In \citetalias{tejeda17} we found that if 
$\Delta \en$ is computed in a simulation as the width of 98 per cent of the 
energy distribution centred on the median value, the $n=0$ approximation 
applies reasonably well for $1 \lesssim \beta \lesssim 4$, while the $n=2$ 
approximation is more suitable for the $\beta \gtrsim 4$ case.

In this paper we will extract $\Delta \en$ as $\sigma_\en$, the standard 
deviation of the energy distribution, as this is more representative of the shape 
of the actual ${\rm d}M/{\rm d}\en$ curve: a more peaked distribution will have 
a smaller  $\sigma_\en$, while the 98 per cent width-definition simply depends 
on the difference in energy between the 1st and 99th percentiles.

\subsection{Periapsis passage}
\subsubsection{Compression and bounce}
As the star approaches periapsis, the fluid elements are compressed in
the vertical direction and the central density of the star can
increase by up to a few orders of magnitude. 
In the case of white dwarfs, the compression can be severe enough to
trigger thermonuclear ignition
\citep{luminet89,rosswog08,rosswog09,anninos18}. In the case of
main-sequence stars, the temperatures are too low and the time scales
too short to make this scenario likely. 

\begin{figure*}
\includegraphics[width=\textwidth]{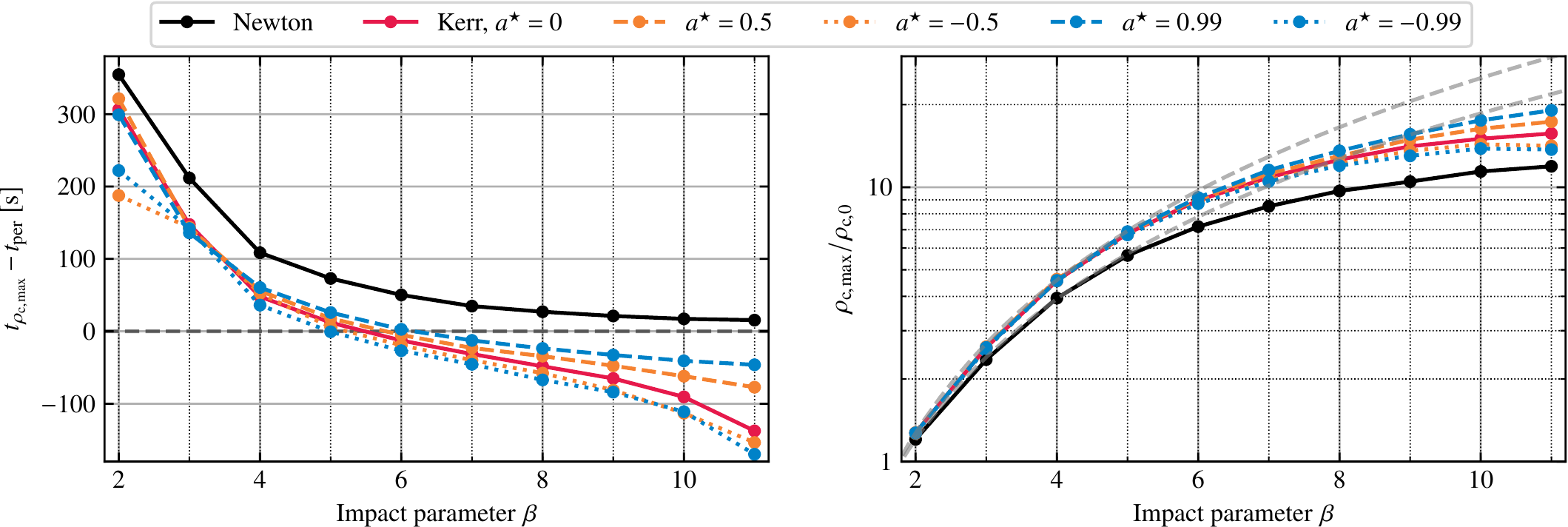}
\caption{\textit{Left panel.} Time when the maximum compression is 
attained, $t_{\rho_{\rm c,max}}$, measured since the time of periapsis 
crossing $t_{\rm per}$, as a function of the penetration factor $\beta$. 
Negative values occur when the star experiences the maximal compression 
before reaching periapsis. \textit{Right panel.} Maximum central density, 
$\rho_{\rm c,max}$, scaled by the initial central density $\rho_0$, as a 
function of the penetration factor $\beta$. The dashed gray lines show 
the $\propto \beta^{1.7}$ and $\propto\beta^{1.85}$ fits, accurate for 
$\beta \lesssim 5$ for Newtonian and relativistic simulations, respectively. 
In this and the rest of the figures in this paper, the colour coding is as follows. 
Newtonian: solid black line; Schwarzschild (Kerr, $\aBH=0$): solid red line; 
Kerr, $\aBH=\pm 0.5$: dashed (dotted) orange line; Kerr, $\aBH=\pm 0.99$: 
dashed (dotted) blue line, as specified in the legend at the top of the figure.}
\label{Fig01}
\end{figure*}

The analytical estimations of \citet{carter83} predicted that for
$\gamma_{\rm ad}=5/3$, the central density and
temperature at the point of maximum compression scale as
$\rho\propto\beta^3$ and $T\propto\beta^2$, respectively. 
Subsequent numerical simulations
have not confirmed this density scaling; starting with the earliest
simulations of \citet{bicknell83}, through the first relativistic
simulations of \citet{laguna93}, and on to the present-day
high-resolution simulations, the scaling found numerically has been
closer to $\rho\propto\beta^{1.5}$, though $T\propto\beta^2$ is, usually, a
fairly reasonable approximation. In our simulations (Fig.~\ref{Fig01}),
we find a scaling of $\rho\propto\beta^{1.7}$ for the Newtonian
simulations, and $\rho\propto\beta^{1.85}$ for the relativistic ones,
although this is only applicable for low $\beta$ ($\lesssim 4$). The
slope becomes significantly milder in strong disruptions, approaching
$\rho\propto\beta^{0.65}$ for the Newtonian simulation and
$\rho\propto\beta^{0.2}$, $\beta^{0.5}$ and $\beta^1$ for the retrograde,
non-spinning, and prograde simulations, respectively. The compression
factor decreases from $\aBH=1$, through $0$, towards $-1$. In our
simulations, the largest ratio which we observe is in the case of
$\aBH=0.99$ vs Newtonian, where the former yields a $\sim 60$ per cent 
higher compression factor for $\beta=11$. 

On the other hand, we find $T\propto \beta^2$ to be an excellent approximation, 
particularly for the Newtonian simulations with $\beta\gtrsim 4$, while being 
slightly milder ($T\propto\beta^{1.5}$) for weaker encounters.

\begin{figure*}
\includegraphics[width=\textwidth]{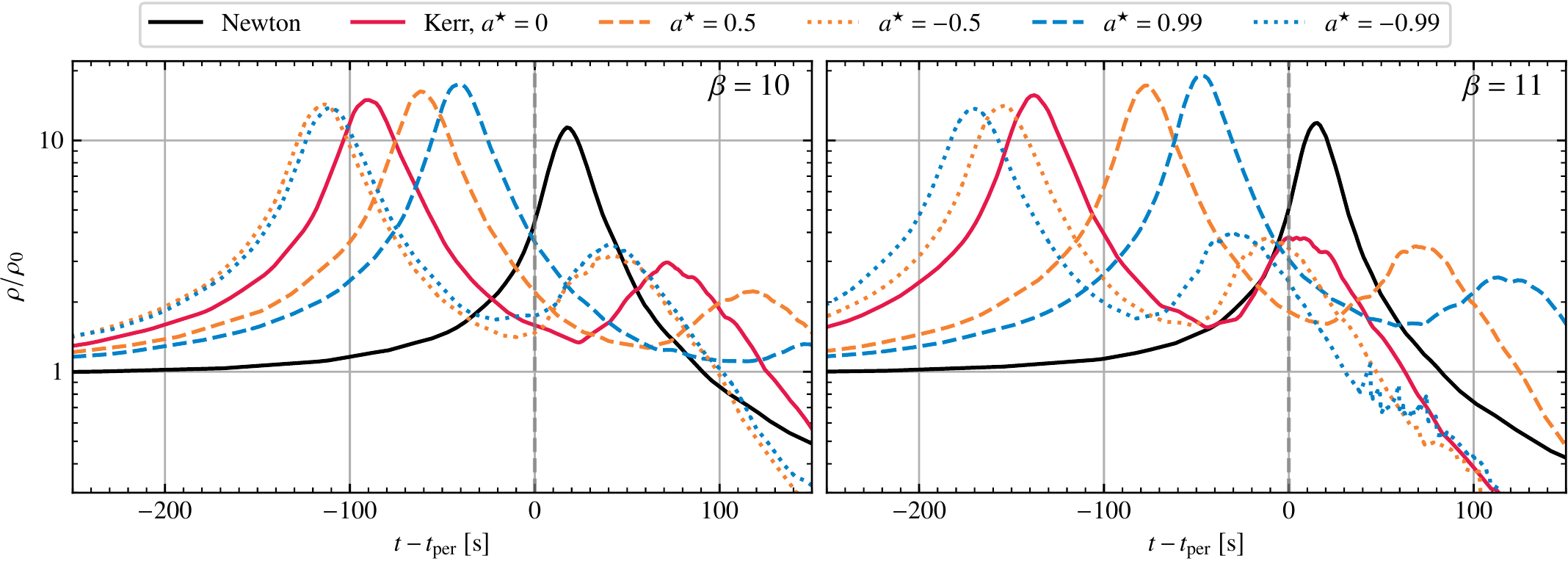}
\caption{Time evolution of the central fluid density scaled by the initial central 
density of the star, $\rho/\rho_0$, for all simulations with penetration factor 
$\beta=10$ (left panel) and $\beta=11$ (right panel).
The plot illustrates how relativistic simulations are able to produce multiple 
bounces at periapsis of which the first (and largest) occurs well before periapsis 
(up to three minutes earlier for $\beta=11$, $\aBH=-0.99$).}
\label{Fig02}
\end{figure*}

\citet{luminet85} also predicted that in the case of relativistic simulations, 
the first ``pancake point'' should be attained before periapsis; for deep 
($\beta\gtrsim 7$) encounters, this leads to a second vertical compression 
and bounce, which is markedly different from the Newtonian case, where 
there is always one unique point of maximum compression, always reached 
just after the periapsis passage.
This feature has been reproduced by \citet{laguna93} and \citet{kobayashi04}, 
and appears in our simulations as well. In Fig.~\ref{Fig02} we show the evolution 
of the central density as a function of time since periapsis, for all of the simulations 
with $\beta=10$. In all five relativistic simulations, the maximum compression 
is attained before periapsis, and there is always a second compression after 
periapsis. The second compression is stronger for the retrograde orbits 
(which have the first bounce earlier; dotted lines) than for the prograde 
orbits (which have the bounce closer to periapsis; dashed lines). 
The $\aBH=0$ case is in between the two.
The typical spacing between the bounces is $\sim$ few minutes. 
Unlike \citet{laguna93}, we do not find evidence for more than two density 
peaks at $\beta=10$, although the second peak is noisier than the first 
(best seen in the red curve of Fig.~\ref{Fig02}), in part due to imperfectly 
captured shock noise, and in part due to the sampling (since the simulation 
does not generate output files at each time step, but only at the synchronization 
points of the individual time steps, the state of the particles in between the output 
times cannot be plotted). On the other hand, for $\beta=11$ we clearly see 
multiple bounces for the retrograde cases (dotted curves).

In the left panel of Fig.~\ref{Fig01} we show the time when the maximum 
compression is attained (measured since periapsis) as a function of $\beta$, 
for all the simulations. It is clear that the Newtonian simulations always have 
the peak after periapsis, sooner (within seconds) for the deepest encounters 
and later (within minutes) for the weakest ones. In the relativistic case, however, 
maximum compression is attained before periapsis for all simulations with 
$\beta \gtrsim 5$, with retrograde orbits reaching it much earlier (up to 
$\sim$ few minutes) than prograde ones.

The multiple bounces are interesting as they may give rise to the formation 
of multiple shocks propagating from the core to the surface. The resulting 
shock breakouts (\citealp{guillochon09}; \citealp{yalinewich19}) may produce 
a distinctive X-ray signature, particularly for negative BH spins, where we see 
more than two compression points in very deep encounters ($\beta > 10$).

\subsubsection{Morphological classes}
\begin{figure*}
\includegraphics[width=\textwidth]{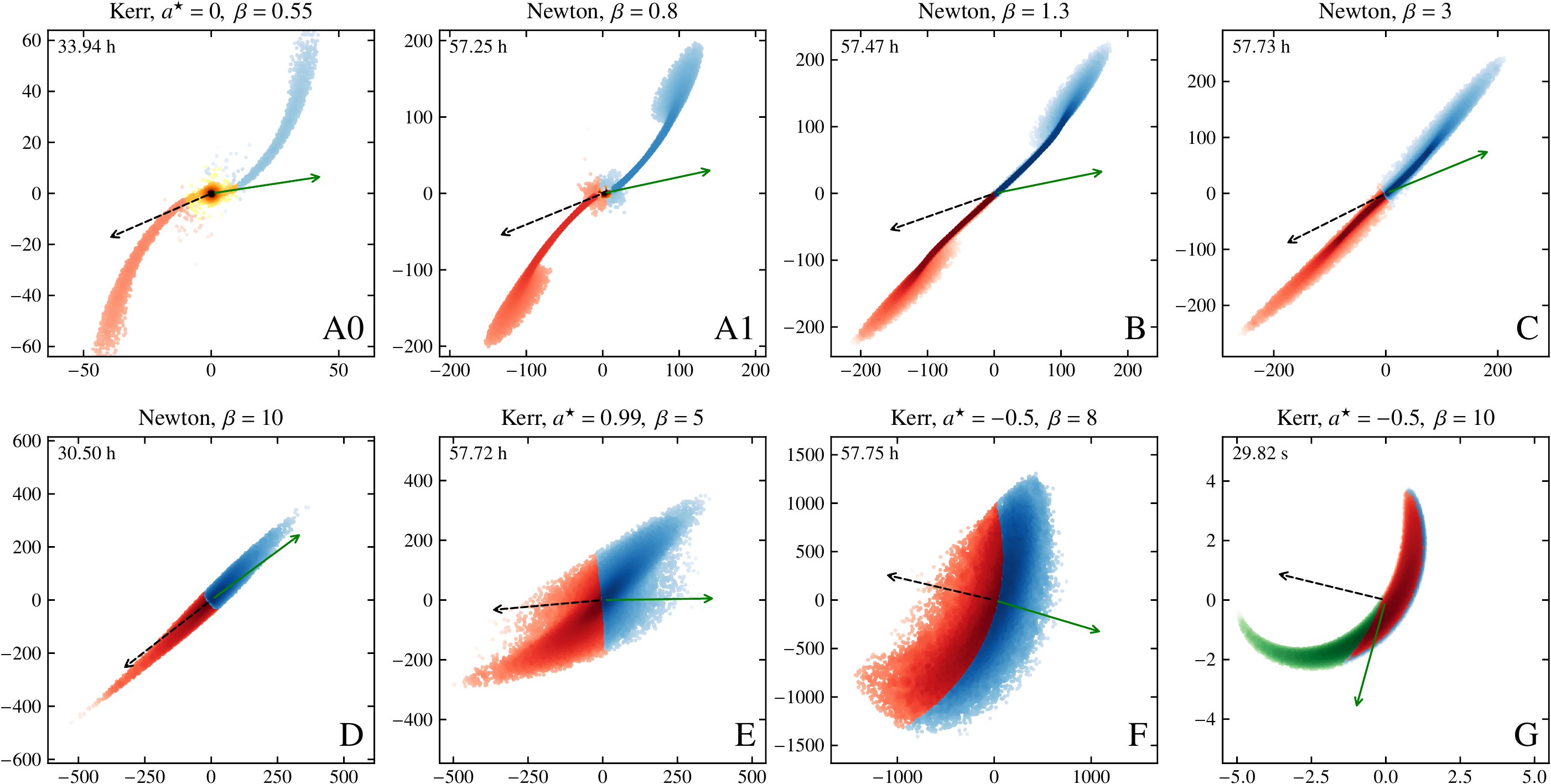}
\caption{Morphological types of debris stream seen in our simulations. The 
colour coding denotes self-bound (yellow), bound (red), unbound (blue) and 
plunging (green) particles, with the colour intensity being related to the logarithm 
of the density (without the colour scale being the same in all pictures). Types 
E, F and G are only seen in relativistic simulations. The axes are transformed 
spatial coordinates $(\tilde{x},\tilde{y})$, as described in the main text, given 
in units of $GM/c^2$ and with the origin in the centre of mass of the debris. 
The details of the model (gravity, spin, impact parameter) are given in the title 
of each panel, the letter describing the morphological type (from A to G, as 
discussed in the main text) is given in the lower right corner, and the physical 
time of the snapshot (measured from the periapsis passage) is given in the upper 
left corner of each panel. The dashed black arrow points in the direction of the 
black hole, while the solid green arrow points in the direction of motion of the 
centre of mass of the stellar debris.}
\label{Fig03}
\end{figure*}

Our simulations produced a large variety of morphological classes for the tidal 
debris stream, some of which have not yet been presented in the literature.  
Based on geometry alone, we find that tidal disruptions may result in 
\textit{seven distinct morphological classes} (see Fig.~\ref{Fig03}):

\begin{description}
\item Class A: a surviving core surrounded by two tidal arms, without (A0) 
or with (A1) tidal lobes at the end ($0.5\lesssim\beta\lesssim 0.9$);

\item Class  B: a thin, well-defined tidal bridge and two well-defined tidal 
lobes at the end; no visible core ($0.9\beta\lesssim 1.5$);

\item Class C: a thick, poorly-defined tidal bridge, with two poorly-defined 
tidal lobes that span most of the length of the bridge ($1.5\beta\lesssim 3$);

\item Class D: a thin, airfoil-shaped stream (Newtonian: $\beta \gtrsim 4$); 
\end{description}
The Kerr cases additionally result in:  
\begin{description}
\item Class E: two nearly-triangular, overlapping tidal lobes with no tidal bridge 
in between (Kerr only: $4\lesssim \beta \lesssim 6$);

\item Class F: a thick stream that is only accreting from its inner part (Kerr only: 
$6\lesssim \beta \lesssim 9$, but highly dependent on the spin);

\item Class G: a spiral that is accreting from its near-end and that is expanding 
ballistically (Kerr only: $\beta \gtrsim 9$, again depending on the spin).
\end{description}

\begin{figure*}
\includegraphics[width=\textwidth]{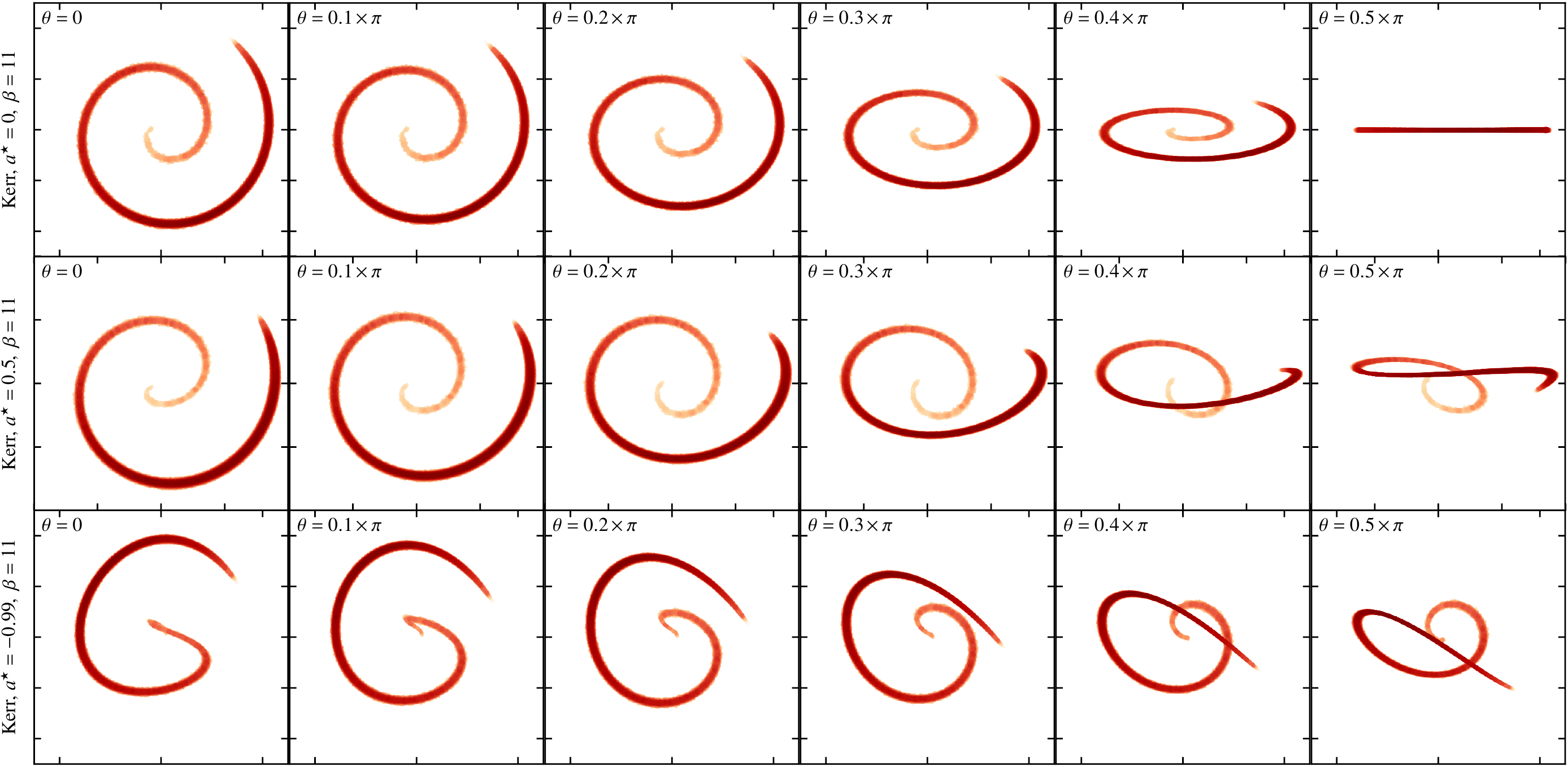}
\caption{
Impact of spin on the long-term evolution of the tidal debris morphology. 
The panels represent three simulations with $\beta=11$ and BH spin parameter 
$\aBH=0$ (\textit{top row}), $0.5$ (\textit{middle row}) and $-0.99$ 
(\textit{bottom row}). The post-disruption snapshots of the three simulations 
correspond to $\sim 7$ minutes  after the first periapsis passage. 
The coordinate system used is $(\tilde{x},\tilde{y})$ as discussed in the main text; 
the six columns represent the same distribution, rotated by a polar angle 
$\theta$ ranging from $0$ (face-on) to $\pi/2$ (edge-on), in order to better 
illustrate the three-dimensional distribution of the debris stream around the spinning BHs.
}
\label{Fig04}
\end{figure*}

At low $\beta$, the Newtonian and relativistic encounters are similar, passing 
progressively through stages A, B, and C; however, in so far as the relativistic 
encounters are more disruptive in terms of the mass removed from the star, 
they reach stages B and C at lower impact parameters. 

After $\beta\sim 2$ ($r_{\rm p}/r_{\rm g}\approx 23.5$), Newtonian and 
relativistic encounters become qualitatively different:
the Newtonian encounters with $\beta \gtrsim 4$ are similar, resulting in 
virtually identical airfoil-shaped debris streams that expand adiabatically. 
For the encounters in Kerr, however, we observe several new morphological 
classes, all of them ultimately linked to the individual relativistic precession 
of the fluid elements: up to $\beta\approx 5$, the tidal tails merge into a single, 
double-triangular shaped stream with no tidal bridge. After that, up until 
$\beta\approx 9$, the debris takes the shape of a very thick, banana-shaped 
stream that accretes from its inner part. Above $\beta \sim 9$, the stream 
becomes a spiral expanding ballistically, with one end ``anchored'' to the BH.

We note that the relativistic debris in panels E and F in
Fig.~\ref{Fig03} exhibits a considerably larger width than in the
Newtonian case, due to the differential periapsis shifts imparted on
the different fluid streams during the periapsis passage. The prospect
of observing such debris streams is promising: the unbound material
keeps expanding and cooling adiabatically, generating an optical
transient from hydrogen recombination  \citep{kasen10}.
It would be plausible to make the assumption that the axis ratio
  $E_t$ of the debris in the orbital plane, in the presence of strong
  periapsis shift, is of order 1, as can be seen in
  classes E and F, instead of $\sim 10$, as was assumed by
  \citet{kasen10}, and which is in agreement with our Newtonian
  simulations represented by class D. In this case, both the expansion
  time $t_e$, defined by \citet{kasen10} in their Eq.~(8) as scaling
  with $\propto E_t^{1/3}$, and the time at which the transient is
  expected to occur, $t_t$, given in their Eq.~(19) with the same
  scaling, would be reduced by a factor of $\sim 2$. In order to test
  this, we extract the times at which the mean and maximum
  temperatures of the debris stream drop below $10^4~{\rm K}$ in two
  simulations with $\beta=6$ (Newtonian and Kerr with
  $\aBH=0$). For the mean temperatures, we find the Newtonian time
  to be $\sim 24$ h, compared to $\sim 8.8$ h for Kerr,
  representing a speed-up of $\sim 2.7$, in agreement with our very
  simple order-of-magnitude analytical estimate. 

If, instead, we consider the maximum temperature, the contrast is much larger: 
in the Newtonian case, the maximum temperature, at the centre of the debris 
stream, only drops below  $10^4~{\rm K}$ after $\sim 160$ days, while in the 
Kerr case it  takes merely $\sim 1.5$ days, representing a speed-up of more than $10^2$. 
In any case, both effects are greatly diminished for $\beta \lesssim 3$, where 
the periapsis shift is not strong enough to generate the $\sim$ 1:1 aspect ratio 
of the debris in the orbital plane.

Another scenario is the production of a $\gamma$-ray afterglow following the 
collision of the expanding debris with molecular clouds  \citep*{chen16}. 
The effect of relativistic periapsis shift is to significantly increase the solid angle 
of the unbound ejecta, reducing the time it takes to end the free expansion and 
begin the Sedov-like phase, as predicted by \citet{khokhlov96} though never 
followed-up with three-dimensional relativistic simulations. The velocities of 
the ejecta are similar in the Newtonian and relativistic simulations (since the parabolic 
velocities are comparable, and of the order of $\sim$ few percent of the speed of light), 
however the expansion velocity (relative to the centre of mass) is higher in the relativistic 
simulations by $\sim$ 50 per cent (below $\beta=1$), 10 per cent (for $1 < \beta < 4$), 
and up to 300 per cent (in deeper encounters); the effect is enhanced for retrograde orbits, 
and diminished for prograde orbits, as compared to the Schwarzschild case. 
This would be expected to significantly enhance the radio signal that is produced once 
the unbound part is braked by the ambient gas, as the total power radiated in 
bremsstrahlung scales with the square of the velocity (\citealp{landau71}).

For case G, the spiral eventually ends up winding multiple times around the BH. 
The spiral shown for class G  is much thinner than the debris stream in classes E and F, 
but note that the time of the snapshot is a mere $\sim$ 30 seconds after the periapsis 
passage,  just before the plunge of the most bound particle into the event horizon, 
as compared with $\sim 57$ hours for E and F. The spiral, however, continues to 
expand because of the differential periapsis shift, and it eventually reaches a 
comparable width to cases E and F (based on ballistic extrapolation). Running the full 
simulation, however, would be problematic, due to the imperative of accurately treating 
the plunge and the second periapsis passage, which is outside the scope of this paper.

We also note that we have found the bound and unbound debris to be mixed (as 
previously observed in the simulations of \citealp{cheng14b}), under the action of 
the different periapsis shifts. This contrasts with  the Newtonian case, where the 
bound and unbound debris are always separated by the initial trajectory of the centre of mass. 
The effect only appears in very close ($r_{\rm p}/r_{\rm g} \lesssim 5$) encounters, 
where a crescent-shape debris stream is formed (as seen before in \citealp{laguna93}; 
\citealp{kobayashi04}; \citealp{cheng14b}; \citetalias{tejeda17}).
Due to the same mixing, a significant part of the plunging material (which is marked 
with green in plot G of Fig.~\ref{Fig03}) may be energetically unbound, invalidating 
the premise (otherwise valid for the Newtonian case) that ``half'' of the debris always 
escapes (see Table~\ref{table:plunge}). Nevertheless, we observe that the ratio of bound 
to unbound plunging material is not 1, but ranges from $\sim 1.4$ (for $\aBH=-0.99$) 
to $\sim 2.3$ (for $\aBH=0.5$). The $a=0.99$ case produces a negligible amount of 
plunging material, since the periapsis is further from the event horizon. The most 
dramatic effect which we see, three-dimensional spirals, appears due to Lense--Thirring 
precession, i.e. only around Kerr BHs with $a \neq 0$; see Fig.~\ref{Fig04}; we have 
first presented such a geometry in Sec. 5.3 of \citetalias{tejeda17}.

\begin{table}
\caption{Percentage of particles on plunging (green in Fig.~\ref{Fig03}), bound (red), 
and unbound (blue) orbits. The bound debris will circularize and give rise to an 
accretion disc of comparable mass.}
\label{table:plunge}
\begin{center}
\begin{tabular}{lclll}
Spin & $\beta$ & Plunging [\%] & Bound [\%] & Unbound [\%] \\
\hline
$-0.99$ & 9 & 0.38 & 47.89  & 51.73 \\
$-0.99$ & 10 & 23.60 & 34.25  & 42.15 \\
$-0.99$ & 11 & 48.52 & 20.57  & 30.91 \\
\hline
$-0.5$ & 9 & 0.26 & 48.01  & 51.73 \\
$-0.5$ & 10 & 23.06 & 34.21  & 42.73 \\
$-0.5$ & 11 & 49.05 & 19.61  & 31.34 \\
\hline
0.0 & 9 & 0.07 & 48.53  & 51.40 \\
0.0 & 10 & 5.95 & 44.66  & 49.39 \\
0.0 & 11 & 38.96 & 25.43  & 35.61 \\
\hline
0.5 & 9 & 0.23 & 48.89  & 50.88 \\
0.5 & 10 & 0.22 & 48.17  & 51.61 \\
0.5 & 11 & 2.83 & 46.87  & 50.30 \\
\hline
0.99 & 9 & 0.31 & 48.99  & 50.70 \\
0.99 & 10 & 0.32 & 48.70  & 50.98 \\
0.99 & 11 & 0.30 & 48.17  & 51.53 \\
\hline
\end{tabular}
\end{center}
\end{table}

We observe that in the Kerr case the debris stream tends to puff up due to 
Lense--Thirring precession, an effect that does not exist in Newtonian simulations. 
This may have implications for how long such a TDE can avoid detection, as the 
general prediction is that a thin-enough stream will avoid self-intersection for 
many orbital periods (\citealp{guillochon15}).

While reviewing how nodal precession may prevent the self-intersection of the 
debris stream, \citet{stone19} pointed out that streams in SPH simulations with 
adiabatic Equations of State (EOS) tend to puff up quickly due to heating from 
internal shocks, and quickly circularize, while streams with isothermal EOS tend 
to remain narrower for a longer time, avoiding circularization for up to 10 orbital 
periods of the most-bound debris ($t_{\rm min}$). Based on the typical temperatures, 
densities and opacities of the bound TDE debris stream, it is unlikely that it could 
be well described by an isothermal EOS, since it is highly opaque to radiation. 
Nevertheless, the concern that SPH simulations tend to produce puffed up TDE 
debris streams is valid, and we would like to address it, since a number of the results 
in this paper originate in the wider debris streams we obtain for relativistic TDEs. 
In our simulations, since we only treat the first stage of the disruption, internal 
shocks only occur during the strong compression experienced during the first 
periapsis passage. In addition, the debris streams we obtain are much narrower 
in the vertical direction than in the orbital plane (with typical ratios between 10 
and 100), and in any case remain much narrower in the Newtonian case than in 
Kerr (with typical ratios $\sim 10$ for classes E and F vs class D, see Fig.~\ref{Fig03}), 
all pointing towards the thickening being a relativistic, rather than hydrodynamic effect.

\begin{figure}
\includegraphics[width=\columnwidth]{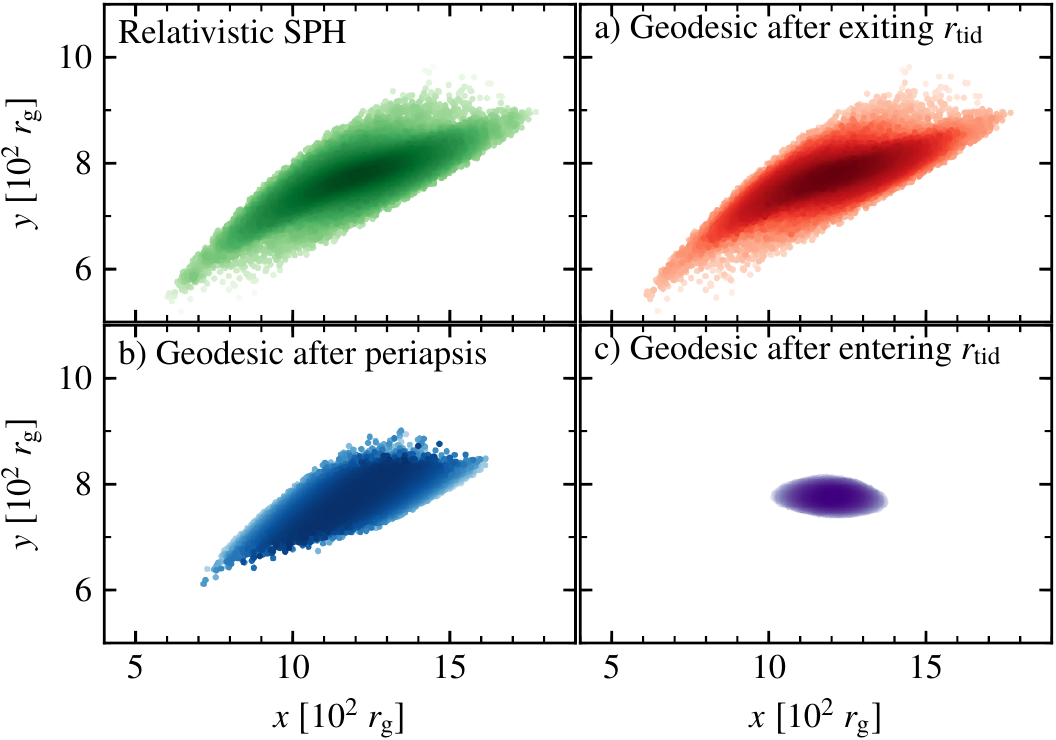}
\caption{The geometry of the debris stream as obtained with a full simulation 
using relativistic SPH (green plot), and by running the first part of the disruption 
with SPH and then extrapolating the geodesic motion assuming the constants of 
motion are frozen-in when: the star exits the tidal radius (red plot), the star passes 
the periapsis (blue plot), or the star enters the tidal radius (purple plot). 
The simulations used a Kerr BH with $\aBH=0.99$, and an impact parameter $\beta=6$. 
The snapshots are all taken at the same time, $\approx 57$ hours after the disruption.}
\label{Fig05}
\end{figure}

Still, in order to test numerically that the puffing up is solely the result of geodesic 
motion, and that  hydrodynamic forces do not affect the stream's evolution (at least 
not before the second periapsis passage), we have also run three control simulations 
of a complete disruption (Kerr,  $\aBH=0.99$, $\beta=6$), by taking a snapshot:
a) as the star exits the tidal radius after disruption,
b) just after the first periapsis passage,
and c) as the star enters the tidal radius before disruption, switching off the self-gravity 
and hydrodynamic forces, and letting the particles evolve on ballistic trajectories alone. 
The results at the end of the simulations (at the same time as the SPH case, 
$\approx 57$ hours after the periapsis passage) are shown in Fig.~\ref{Fig05}. 
We observe that cases a) and b) yield similar results, but only case a) is virtually identical 
to the original simulation, showing that the constants of motion do evolve for some 
time after the bounce, but settle in by the time the star exits the tidal radius. 
The case c) utterly fails to reproduce the geometry of the debris stream, proving 
that the periapsis passage is crucial in determining how the energy and angular 
momentum are redistributed, and so the frozen-in model cannot be applied when 
entering $r_{\rm tid}$ to determine the stream geometry, at least for deep encounters. 
The results also show that the expansion of the debris stream is due to geodesic motion 
alone, as even if the constants of motion are frozen at periapsis, the resulting 
debris stream has a comparable thickness to the one from the full simulation, 
and much thicker than in the Newtonian case.

\begin{figure*}
\includegraphics[width=\textwidth]{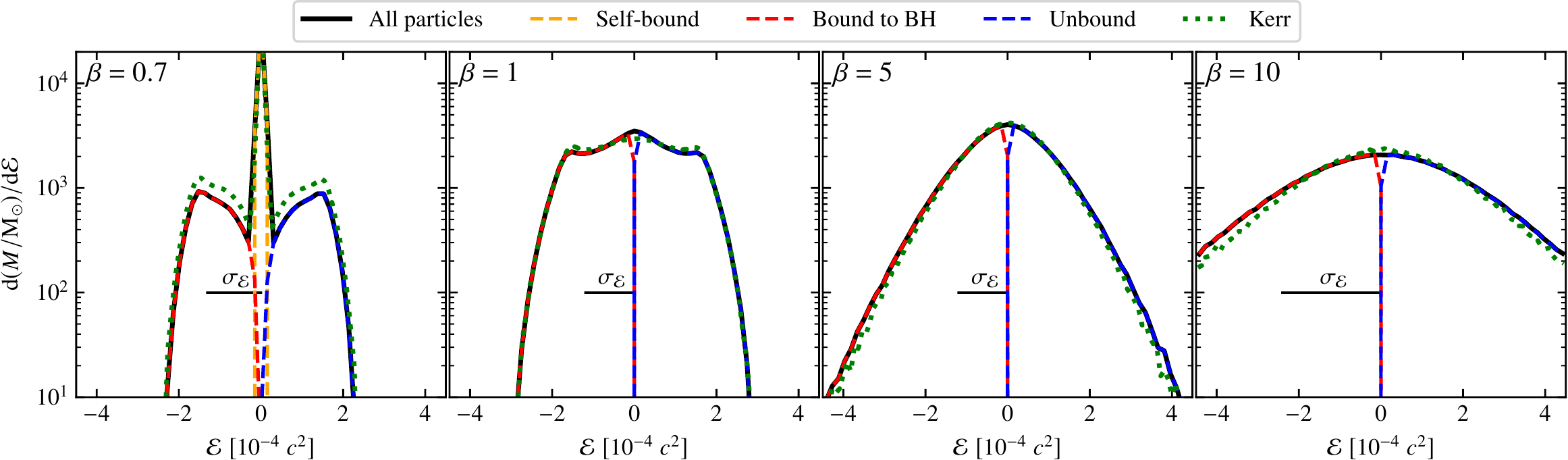}
\caption{Histograms of the specific mechanical energy $\en$ for Newtonian encounters 
with various impact parameters $\beta$. The black line shows the full histogram, 
while the red, blue, and orange lines show the histograms of the bound, unbound, 
and self-bound debris, respectively. The standard deviation $\sigma_{\en}$ is 
computed for all the particles that are not self-bound (i.e., from the red and blue lines).
The mass distribution of energy is well-approximated by a flat line in the vicinity of 
$\en=0$  only between $\beta=1$ and $2$; for $\beta \lesssim 1$, the bound energy 
distribution of the tidal tails has the peak around $\sim\sigma_{\en}$, and drops 
abruptly towards $\en=0$ (predicting a much steeper decay than $\propto t^{-5/3}$); 
for $\beta \gtrsim 4$, the logarithmic distribution resembles a parabola rather than being flat.
To give an idea of the importance of the GR correction, we overplot  the same histogram 
for the Kerr, $\aBH=0.99$ case (dashed green line). The difference is most noticeable 
for the lowest and the highest impact parameters.
}
\label{Fig06}
\end{figure*}

\begin{figure}
\includegraphics[width=\columnwidth]{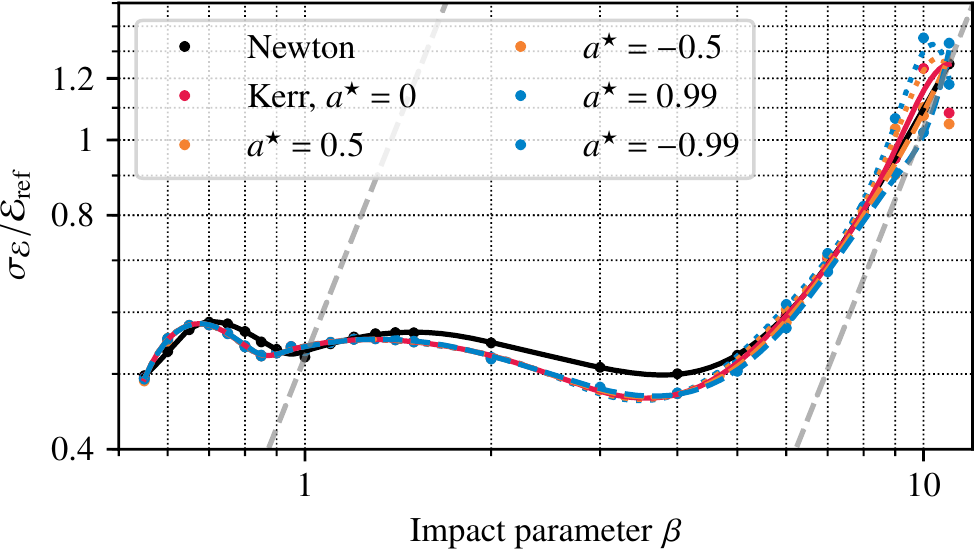}
\caption{Spread of orbital mechanical energies $\en$, calculated as the standard 
deviation of the energies of all SPH particles bound ($\en<0$) and unbound ($\en>0$) 
to the black hole, without including the self-bound particles in encounters with a 
surviving core ($\beta \lesssim 0.9$).
The lines show the $H_{5/3}$ spline fits from Appendix~\ref{appendixA}, while the 
points show the data from our simulations.}
\label{Fig07}
\end{figure}

\subsubsection{Energy spread}
We find that the energy distribution (${\rm d}M/{\rm d}\en$) is not flat around 
$\en=0$ except for a narrow range of impact parameters around $\beta\sim 1$ 
(Fig.~\ref{Fig06}), when most of the matter resides in the thin and dense tidal bridge 
(class C in Fig.~\ref{Fig03}).
For weaker encounters, when the core of the star survives, ${\rm d}M/{\rm d}\en$ 
exhibits broad wings that may evolve at late times under the gravitational influence 
of the self-bound core; for strong disruptions, above $\beta \sim 4$, the logarithmic 
histogram of ${\rm d}M/{\rm d}\en$ can be fitted remarkably well by a generalized 
Gaussian function, with the Gaussian parameters $a$ and $b$ being smooth 
functions of $\beta$, as shown in Appendix~\ref{appendixA}.

The spread in orbital mechanical energies, calculated as the standard deviation of 
the energy distribution, $\sigma_\en$, exhibits little variation with $\beta$ until 
after $\beta \sim4$, where it starts approaching the theoretical predictions of the 
standard frozen-in model, $\sigma_\en\propto \beta^2$ (Fig.~\ref{Fig07}). 
These results are somewhat in contradiction with those recently presented by 
\citet{steinberg19}, who found that above $\beta\sim 5$ the spread in energy is 
nearly insensitive with $\beta$. Apart from using very different codes and computing 
the energy spread in different ways, it is difficult to understand well the origin of this 
difference, as they do not present histograms of the energy distribution. 
We strongly emphasize that the energy spread, in itself, does not offer much information 
about the disruption, in general, or the fallback rate, in particular, unless it is coupled 
with the (quite erroneous) assumption that the energy distribution is flat, which only 
holds around $\beta\sim 1$.

\subsection{Fallback}
\begin{figure*}
\includegraphics[width=\textwidth]{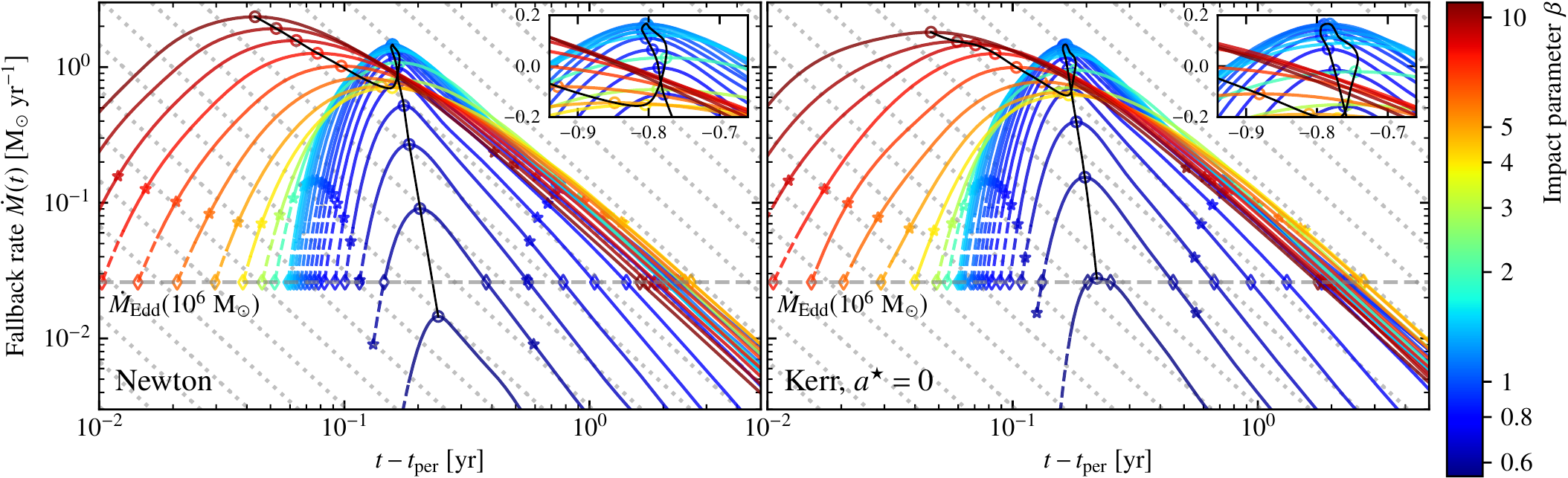}
\caption{Mass fallback rate $\dot{M}(t)$ as a function of the time elapsed since the 
first periapsis passage for the Newtonian case (\textit{left panel}) and for the Kerr 
case with spin parameter $\aBH=0.99$ (\textit{right panel}). The various colours 
correspond to different impact parameters $\beta$, from 0.55 (dark blue) to 11 
(dark red). The symbols mark the times when the fallback rates are equal to: the 
peak rate $\dot{M}_{\rm peak}$ (open circle), 10 per cent of the $\dot{M}_{\rm peak}$ 
(star), and the Eddington limit for a $10^6~{\rm M}_{\sun}$ BH assuming a radiative 
efficiency $\epsilon=0.1$ (open diamond). The diagonal, dotted gray lines show 
the slope of the $t^{-5/3}$ power law decay, while the horizontal, solid gray line 
marks the fallback rate corresponding to the Eddington limit. The inset shows a 
zoom on the region around $\dot{M}_{\rm peak}$ for $\beta=1$, similar to Figure 5 
in \citetalias{guillochon13}. The dashed portions at the beginning of some of the 
curves show extrapolated data where the resolution of the most bound debris was 
too low to extract a meaningful histogram. The data are binned logarithmically and 
smoothed using a spline fit, as detailed in Appendix~\ref{appendixB}.
}
\label{Fig08}
\end{figure*}

In Fig.~\ref{Fig08} we present the fallback rates, $\dot{M}(t)$, for the Newtonian 
simulations (left panel) and for the Kerr case with $\aBH=0.5$ (right panel). The 
procedure for binning the data is discussed at length in Appendix~\ref{appendixB}. 
The log--log plot is similar to the one presented in Fig.~5 of \citetalias{guillochon13}, 
although the parameter range is now extended to $\beta=11$. The fact that, up to 
$\beta\sim 2$, the results match so well the ones from the reference paper is a 
non-trivial test of both, since the two sets of simulations have been performed with 
different codes, using different formalisms (high-resolution, grid-based simulations, 
with a multipole gravity solver, in the rest frame of the star, vs medium-resolution, global, 
tree-based SPH particle simulation), different ways of setting up the initial conditions 
and of postprocessing the data, etc. We even reproduce the feature of 
$\dot{M}_{\rm peak}$ discovered by \citetalias{guillochon13} at around $\beta\sim 1$, 
where the initial trend at low $\beta$, towards earlier and higher peaks with 
increasing $\beta$, reverses to later and lower ones. We find, however, that the 
trend reverses again around $\beta\sim 3$, where the peak starts shifting to 
significantly higher accretion rates and to earlier times. Our explanation for this 
behaviour is related to the occurrence of shocks during the periapsis passage, 
which does not happen at lower $\beta$, as will be discussed later on.

\begin{figure*}
\includegraphics[width=\textwidth]{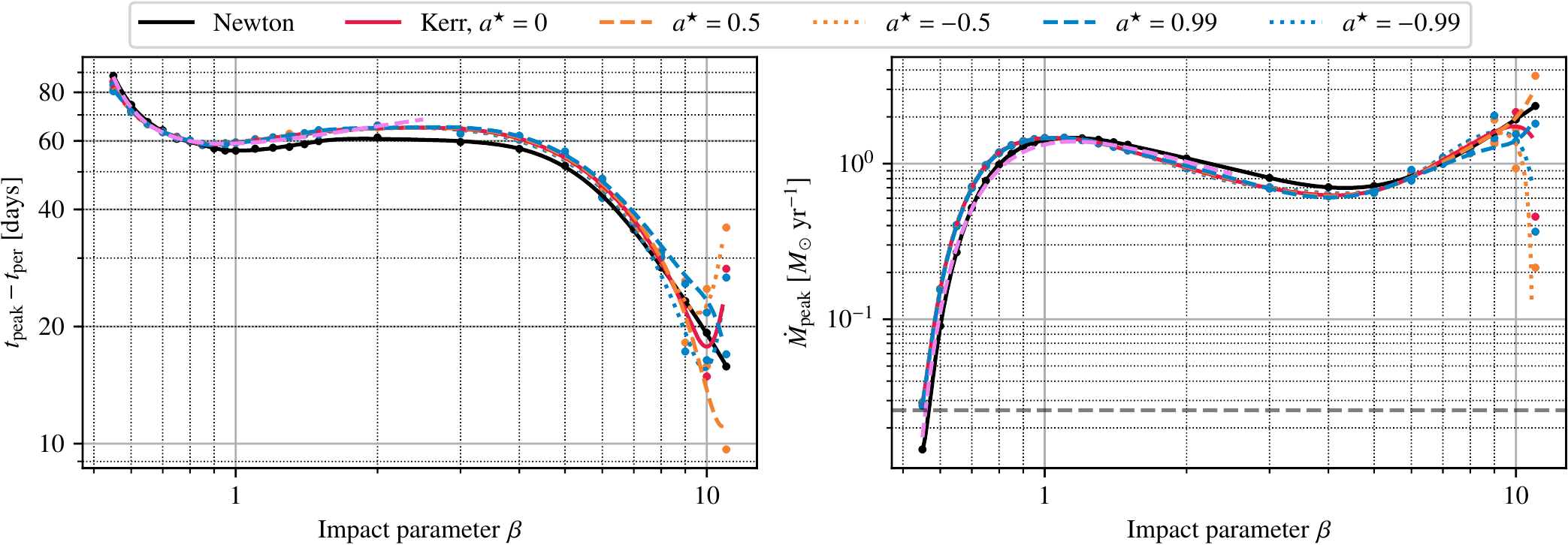}
\caption{\textit{Left panel.} Time when the maximum fallback rate is achieved, 
measured in days elapsed since the first periapsis passage. \textit{Right panel.} 
Peak mass fallback rate, $\dot{M}_{\rm peak}$. The horizontal dashed gray line 
shows the fallback rate corresponding to the Eddington luminosity for a 
$10^6~{\rm M}_{\sun}$ BH, assuming a radiative efficiency $\epsilon=0.1$. 
The lines show the $A_{5/3}$ and $B_{5/3}$ spline fits from Appendix~\ref{appendixA}, 
while the points show the data from our simulations, extracted as described in 
Appendix~\ref{appendixB}; the fits from \citetalias{guillochon13} are overplotted 
with a dashed, purple line.}
\label{Fig09}
\end{figure*}

In Fig.~\ref{Fig09} we present the times and magnitudes of the peak fallback rate, 
$t_{\rm peak}$ and $\dot{M}_{\rm peak}$. For $\beta < 2$, the results for the 
Newtonian simulations are in agreement with Fig.~12 of \citetalias{guillochon13}, 
whose fit curves are overplotted with a dashed purple line. Our results also agree 
with the $\beta=1$ tidal disruptions of \citet{cheng14b}, who concluded that 
Newtonian rates have a slightly earlier rise, while GR rates exhibit: a more gradual 
rise, a higher peak, and a later rise above the Eddington limit.
However, the trend reverses after $\beta>4$, with the peak fallback increasing 
and occurring at earlier times.
We find $\dot{M}_{\rm peak}$ to be significantly higher and $t_{\rm peak}$ to 
be earlier than the predictions of the frozen-in model, consistent with the recent 
findings of \citet*{wu18}.

\begin{figure}
\includegraphics[width=\columnwidth]{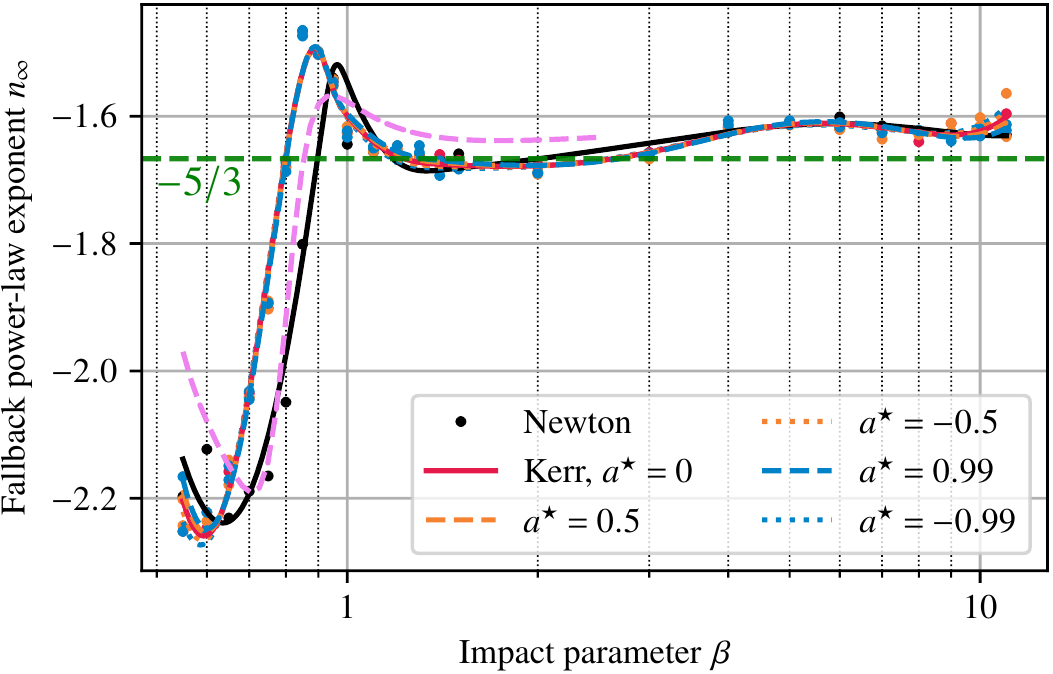}
\caption{Power law index of fallback rate at late times (after it becomes 
sub-Eddington or 10 per cent of the peak rate, whichever occurs later (generally, 
this happens $\sim$ few years after the disruption for a $10^6~{\rm M}_{\sun}$ 
black hole) as a function of the penetration factor $\beta$. 
The lines show the $D_{5/3}$ spline fits from Appendix~\ref{appendixA}, while 
the points show the data from our simulations, extracted as described in 
Appendix~\ref{appendixB}; the fits from \citetalias{guillochon13} are 
overplotted with a dashed, purple line.}
\label{Fig10}
\end{figure}

In Fig.~\ref{Fig10} we plot the slope of the fallback rate at late times. Again, 
the data are fairly similar to what \citetalias{guillochon13} presented in their 
Fig.~7. At high $\beta$, the trend does not disappear, and the slope  remains 
very close to $-5/3$.

\begin{figure*}
\includegraphics[width=\textwidth]{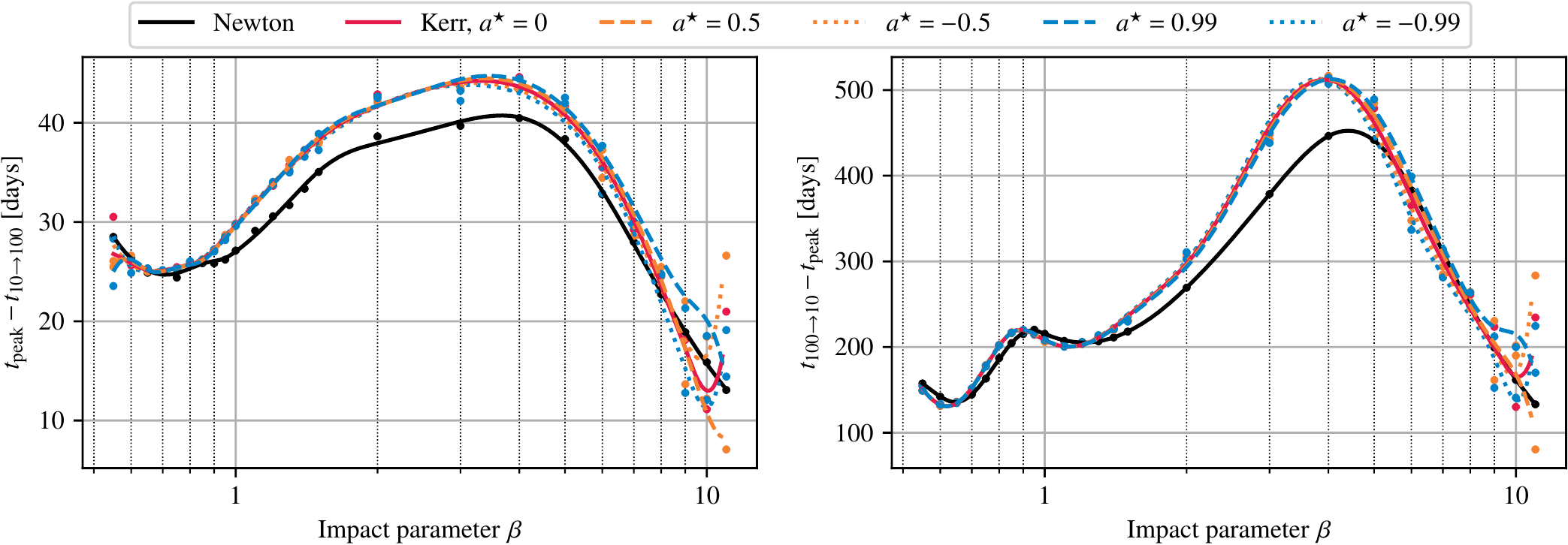}
\caption{\textit{Left panel.} Time of rise from 10 per cent of $\dot{M}_{\rm peak}$ 
to $\dot{M}_{\rm peak}$,  as a function of the penetration factor $\beta$. 
The relativistic curves take up to a couple of days longer to reach the peak. 
\textit{Right panel.} Time of decay from $\dot{M}_{\rm peak}$ to 10 per cent 
of $\dot{M}_{\rm peak}$, as a function of $\beta$. The relativistic fallback rate 
is flatter, most notably around $\beta\sim 4$, where it can last up to $\sim$ 
2 months longer than in the Newtonian case.
The lines show the $E_{5/3}$ and $F_{5/3}$ spline fits from 
Appendix~\ref{appendixA}, while the points show the data from our simulations, 
extracted as described in Appendix~\ref{appendixB}.}
\label{Fig11}
\end{figure*}

In Fig.~\ref{Fig11}, we plot the time of rise from 10 to 100 per cent of 
$\dot{M}_{\rm peak}$ ($\tup$, \textit{left panel}) and the time of decay 
from 100 to 10 per cent of $\dot{M}_{\rm peak}$ ($\tdown$, \textit{right panel}). 
These quantities, although not customarily presented in the literature on 
numerical TDEs, may very much be of interest for the analysis of observational 
data, as they are a good representation of how broad the fallback curves are, 
and of how quickly they rise and fall. Here is where we find the biggest 
relativistic effects, most noticeable at moderate $\beta$ (between $\sim 1$ and 
5). There, the relativistic fallback rates take $\sim$ few days longer to reach 
the peak from $10$ per cent of its value, and significantly longer 
($\sim$ few months) to  decay.

\begin{figure}
\includegraphics[width=\columnwidth]{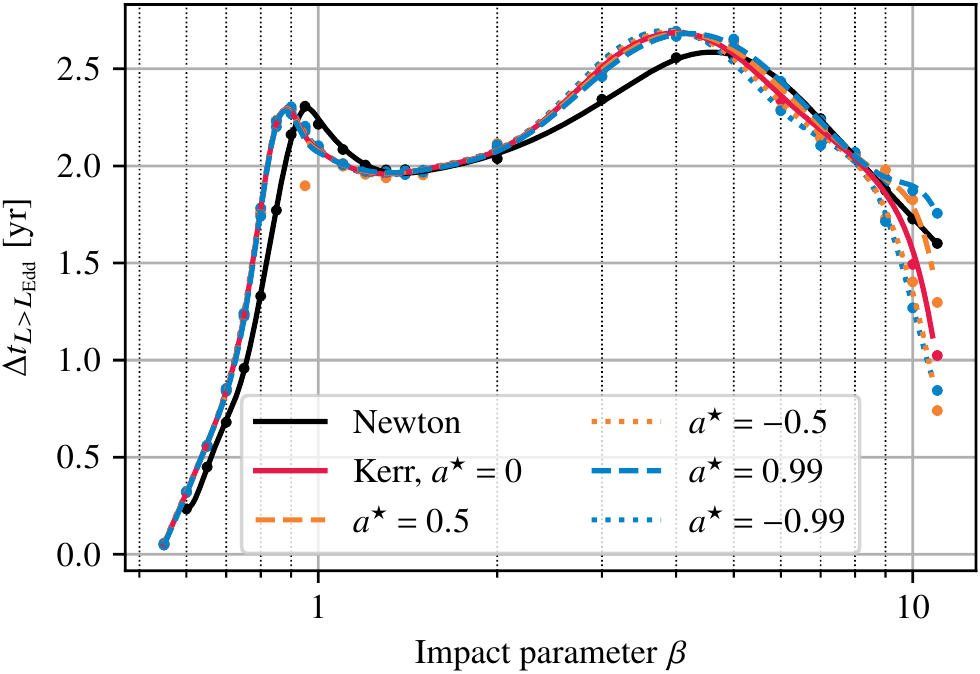}
\caption{Duration of super-Eddington fallback assuming a radiative efficiency 
of $\epsilon=0.1$ and the Eddington limit  $L_{\rm Edd}=4\pi GMc/\kappa$, 
with $\kappa=0.34$~cm$^2$~g$^{-1}$ being the Thomson opacity assuming 
solar abundances. The relativistic simulations result in a slightly longer duration 
of super-Eddington fallback, except for the highest $\beta$, where a significant 
percentage of the stellar material is initially captured on plunging orbits.
The lines show the $G_{5/3}$ spline fits from Appendix~\ref{appendixA}, while 
the points show the data from our simulations, extracted as described in 
Appendix~\ref{appendixB}.}
\label{Fig12}
\end{figure}

In Fig.~\ref{Fig12}, we plot the duration of super-Eddington fallback rate. 
This is not as helpful a quantity as $\tup$ and $\tdown$, since it depends 
non-trivially on the combination of the BH mass $M$ and the shape of the 
fallback rate curve. The trends that we find for our disruptions by a 
$10^6~{\rm M}_{\sun}$ BH are that: a) relativistic super-Eddington fallback 
lasts from $\sim$ few months (for $\beta <1$) to $\sim 2$ years (for 
larger $\beta$) longer than in the Newtonian case; b) the duration of 
super-Eddington fallback is severely reduced in the high-$\beta$ relativistic 
encounters, due to the large amount of material that plunges directly into 
the BH; c) for moderate disruptions, the duration in relativistic
encounters may be longer by $\sim$ a month than in the Newtonian case.

To date, the most comprehensive numerical investigation of the TDE fallback 
process across a wide range of impact parameters has been undertaken in 
\citetalias{guillochon13}, in whose Appendix A the authors present fitting 
parameters for $\dot{M}_{\rm peak}$, $t_{\rm peak}$, $\Delta M$ and 
$n_\infty$, as functions of $\beta$, scaled with $M$, $m_\star$ and 
$r_\star$, and with separate fittings for $\gamma=4/3$ and $\gamma=5/3$. 
These fits have proven invaluable in subsequent detections of TDE flares 
(e.g., \citealp{gezari15}; \citealp{leloudas16}), and have helped narrow 
down the parameter space of those events. In Appendix~\ref{appendixA} 
we present an extended range of fit formulae, based on the simulations 
in this paper, which extends the ones given in \citetalias{guillochon13} 
to larger $\beta$ (up to 11) and to rotating black holes.

\section{Discussion}
A comparative look at most of the plots in this paper reveals that,
 depending on the impact parameter $\beta$, tidal disruptions fall into 
 three categories (we will illustrate with values of $\beta$ corresponding 
 to $\gamma=5/3$; note that the exact values of $\beta$ are different 
 for other equations of states, most notably for polytropes with 
 $\gamma=4/3$, although we expect the trend itself to be unchanged). 

\subsection{Weak disruptions}
\begin{figure}
\includegraphics[width=\columnwidth]{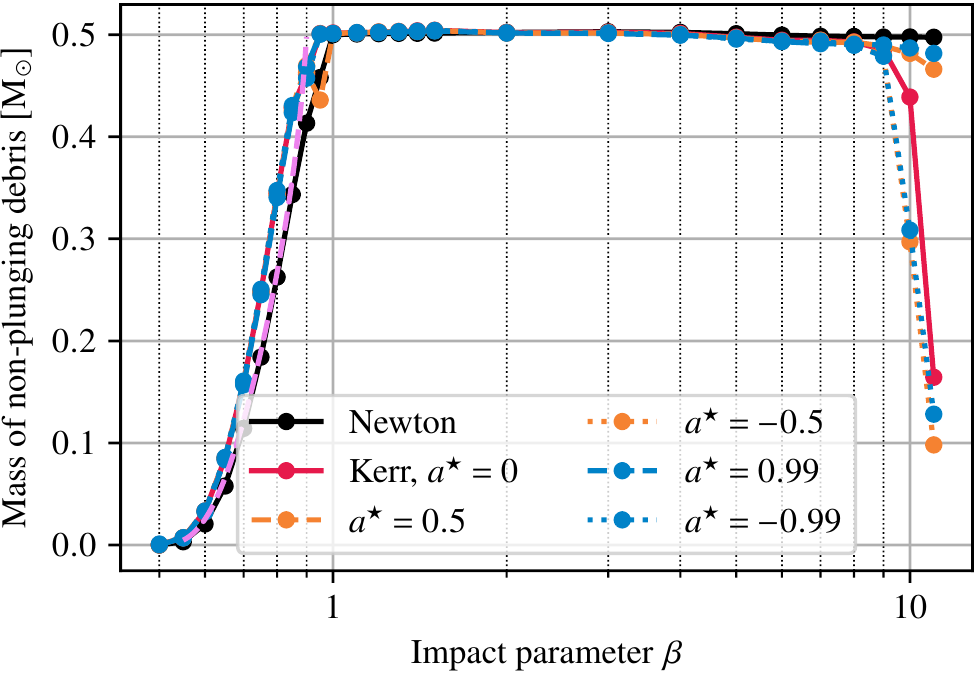}
\caption{Mass fraction of material removed from the star and bound 
to the black hole ($\en<0$) without being on plunging orbits. This is 
the material that will eventually circularize and form an accretion disc. 
Values given in units of ${\rm M}_{\sun}$, as a function of the penetration 
factor $\beta$.
The fits from the Appendix of \citetalias{guillochon13} are overplotted 
with a dashed, purple line.
}
\label{Fig13}
\end{figure}

At low $\beta$ ($\beta \lesssim 0.9$), the star suffers a partial disruption.
The mass fraction $\Delta M/m_\star$ removed from the star and bound 
to the black hole varies greatly, ranging from $\sim 0$ around $\beta=0.5$ 
to $\sim  0.5$ at the disruption limit $\beta_{\rm d}\sim 0.9$ (Fig.~\ref{Fig13}; 
the exact value of $\beta_{\rm d}$ greatly depends on $\gamma$, see 
\citetalias{guillochon13}); in the relativistic disruptions, more mass is 
stripped from the core than in the Newtonian case, from $\gtrsim 100$ 
per cent more (at $\beta=0.55$) to $\sim 10$ per cent more (at 
$\beta=0.9$), in agreement with the pseudo-relativistic simulations of 
\citet[Fig.~3; note that the effect is greatly exacerbated for larger BH 
masses]{gafton15}. The effect of the BH spin is small but consistent, 
with $\sim 1$ per cent less mass in the surviving self-bound core for 
$\aBH = -0.99$ and about $\sim 1$ per cent more mass in the 
surviving self-bound core for $\aBH = 0.99$ as compared to $\aBH=0$.

The morphology of the tidal debris consists of a surviving core and two 
tidal tails (A0 and A1 in Fig.~\ref{Fig03});
the time to peak fallback rate is long, of the order of $\sim 2$--3 months, 
and the fallback rate varies significantly, from $\sim L_{\rm edd}$ at 
$\beta\sim 0.55$ to $\sim 10^2~L_{\rm edd}$ at $\beta\sim 9$ 
(Fig.~\ref{Fig09}); the relativistic simulations yield up to twice the 
fallback rate for the lower range of $\beta$.
The durations of the rise and decay of the fallback rate (quantified 
as the time it takes to get to 10 per cent of $\dot{M}_{\rm peak}$ to 
$\dot{M}_{\rm peak}$, and back) are long and scale inversely with 
$\beta$ (Fig.~\ref{Fig11}).
The duration of super-Eddington flow is highly variable (Fig.~\ref{Fig08}), 
from $\sim 2$ months for $\beta\sim 0.6$ to $\sim 2$ years for 
$\beta\sim 0.9$.
The energy spread is relatively high; the energy distribution consists 
only of the tidal tails, which are roughly centred around $\sigma_\en$ 
(Fig.~\ref{Fig06}) and quickly drop towards $\en\sim 0$, predicting a 
sudden drop in the fallback curve.
The long-term fallback exponent, $n_\infty$, is significantly steeper 
than $t^{-5/3}$ for very weak encounters ($n_\infty\propto t^{-2.2}$), 
and milder than $t^{-5/3}$ close to the disruption limit 
($n_\infty\propto t^{-1.5}$) (Fig.~\ref{Fig10}); the reason 
for this is connected to the influence of the surviving core, and 
was explained at length in \citetalias{guillochon13}. We find the 
relativistic trend very similar, although $n_\infty$ is shifted to 
lower $\beta$ as compared to the Newtonian case.

\subsection{Moderate disruptions}
At moderate $\beta$ ($1 \lesssim \beta \lesssim 3$), the star is 
disrupted completely, but {remnants of  self-gravitating cores 
remain: morphologically, in the form of a thin tidal bridge, and 
energetically, in the form of a flat central distribution and 
``wings'' in the energy histogram.
The tidal bridge is kept narrow by self-gravity, but at the two 
ends it broadens into two tidal tails with or without lobes 
(B and C in Fig.~\ref{Fig03}).}
The time to peak is shorter than for very weak disruptions 
($\sim$ 60 days), and the peak fallback rate is higher (by 
about one order of magnitude), but the trend reverses 
around $\beta=1$, as first noted in \citetalias{guillochon13}, 
and also visible in the inset plots in Fig.~\ref{Fig08}); from 
that point, the fallback rate decreases and the time to peak 
increases with $\beta$, reaching a minimum around $\beta=4$ 
(corresponding to the narrowest point in the ${\rm d}M/{\rm d}\en$ 
spread in Fig.~\ref{Fig10}).
The times of the rise and decay of the $\dot{M}$ curve are very 
large, reaching a maximum around $\beta=4$ ($\sim$ 45 days 
for the rise and $\sim$ 500 days for the decay); these encounters 
last the longest (as measured from $\tup$ to $\tdown$), with 
relativistic effects prolonging them even further, by up to $20$ 
per cent around $\beta=4$.
The duration of super-Eddington flow is also  very long, surpassing 
2.5 years for rotating black holes with $M=10^6~{\rm M}_{\sun}$.
The energy spread is small, with the lowest point occurring around 
$\beta=4$; the energy histogram consists of a relatively flat plateau 
of width $\sim\sigma_{\en}$ where most of the mass is located 
(the tidal bridge), and has very little mass outside of it (in the tidal tails).
In this regime, the long-term fallback rate slope settles to the 
canonical $t^{-5/3}$ and exhibits virtually no variation with $\beta$.

\subsection{Strong disruptions}
At high $\beta$ ($\beta \gtrsim 4$), the star is fully disrupted; 
for Newtonian encounters, the morphology is a long stream, devoid of
tidal tails or any other distinguishing features (D in Fig.~\ref{Fig03});
for the Kerr case, the debris is much thicker, with (E) or without (F) 
features reminiscent of the tidal tails; for non-equatorial orbits, the 
nodal precession creates three-dimensional spiral-like debris 
distributions (G) that move ballistically; due to the different shifts 
imparted at periapsis, the resulting structure is expected to thicken 
significantly in time.

The time to peak is very short (of the order of $\sim$ few weeks), 
and the fallback rate is very high, surpassing $2~{\rm M}_{\sun}~{\rm yr}^{-1}$ 
for the deepest encounters; this, however, is not the case in very close 
relativistic encounters, where a significant fraction of the star may 
plunge into a black hole and be promptly accreted (see 
Table~\ref{table:plunge}), leaving less material to fall back and circularize.
The energy spread is large, and it shows a power law dependence 
on $\beta$, only slightly milder than the $\beta^2$ predicted by 
the original frozen-in theory; the logarithmic energy distribution 
assumes a nearly Gaussian shape, whose parameters can be well 
characterized in terms of $\beta$ alone.
The fallback exponent remains very close to $t^{-5/3}$.

\subsection[What happens at beta = 4?]{What happens at $\beta\approx4$?}
We believe that the qualitative change in the disruption around 
$\beta\approx 4$ is related to the shock that forms due to strong 
compression: at $\beta \lesssim 1$, the star is not compressed 
at all, and it is disrupted smoothly, due to the resonant quadrupole 
oscillations induced by the tidal field of the BH; there is no real 
bounce wave to steepen into a shock, because the fluid pressure 
has enough time to continuously react to the (mild) geodesic 
compression during the periapsis passage. At very large $\beta$,  
however, the strong compression is halted by a shock, which 
then rebounds and travels to the surface of the star, as clearly 
confirmed in previous simulations (e.g., \citealp{guillochon09} 
for $\beta=7$ and $\gamma=4/3$). It follows, then, that there 
must be a boundary, in between full disruptions without 
compression or shock (at $\beta=1$) and full disruptions 
with large compression followed by a shock (which we know 
happens at $\beta=7$). This has been hinted at in Sec.~5.2 of 
\citetalias{tejeda17}, but not fully discussed.
Note also that \citet{rosswog09} found a threshold value for 
detonating white dwarfs around $\beta_{\rm crit}\approx 3$, 
which would be consistent with the above interpretation.

\begin{figure*}
\centering
\includegraphics[width=\textwidth]{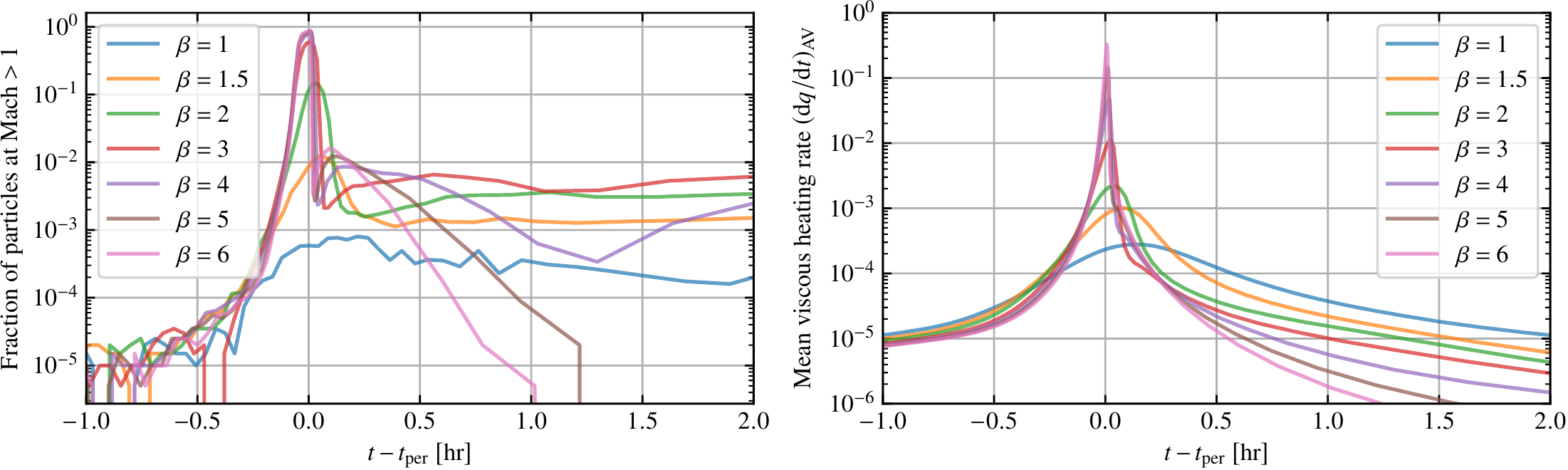}
\caption{Indicators of the presence of shocks in several Newtonian SPH 
simulations, from $\beta=1$ to $\beta=6$. (\textit{Left panel}) Fraction 
of particles having a Mach number larger than 1, as computed with the 
geometrical shock-detection method of \citet{beck16}, as a function of 
the time elapsed since periapsis. (\textit{Right panel}) Mean SPH viscous 
heating rate $({\rm d}q/{\rm d}t)_{\rm AV}$, which is a good proxy for 
the entropy generated in the shock, as a function of the time elapsed 
since periapsis.}
\label{Fig14}
\end{figure*}

We look for the presence of shocks in two ways: first, with the 
geometrical shock-detection method for SPH presented by \citet*{beck16}. 
In Fig.~\ref{Fig14} (\textit{left panel}) we plot the fraction of SPH particles 
that have a Mach speed larger than 1, as computed with the formalism 
described in Section 2 of that paper, as a function of time elapsed 
since periapsis, for seven Newtonian simulations ranging from 
$\beta=1$ to $\beta=6$. We see that for $\beta\gtrsim 3$, virtually 
the entire star is shocked right after the periapsis passage (with 
the curves being nearly identical for all $\beta\gtrsim 4$), while 
for $\beta=1$ less than 0.1 per cent of the star experiences Mach 
$> 1$ at periapsis. The actual location of the shock also differs: 
at low $\beta$, the shock occurs in the tidal tails (as discussed 
before by \citealp{lodato09}),  and the fraction of shocked particles 
is maintained at a level above $0.1$ per cent long after disruption 
($\sim$ few hours), with additional shocks occurring as a result of 
tidally-induced, non-linear stellar pulsations. On the other hand, 
at high $\beta$ ($> 4$), the shock occurs throughout the entire 
star during the first periapsis passage; as the debris stream is now 
devoid of tidal tails and is merely expanding homologously, the 
fraction of shocked particles quickly drops by several orders of 
magnitude within an hour after disruption.

We also analyse the viscous heating rate $({\rm d}q/{\rm d}t)_{\rm AV}$, 
representing the contribution of the SPH artificial viscosity to the 
energy equation, which is a proxy for the entropy generated in the shock. 
In Fig.~\ref{Fig14} (\textit{right panel}) we plot the average  
$({\rm d}q/{\rm d}t)_{\rm AV}$, calculated for all SPH particles 
in the simulations with $\beta=1$ to $\beta=6$, as a function 
of time elapsed since periapsis. 
We see that for the $\beta=6$ case, the average $({\rm d}q/{\rm d}t)_{\rm AV}$ 
is three orders of magnitude larger than for $\beta=1$, confirming 
the generation of significantly more shock entropy for the deeper 
encounters. The rise of the $({\rm d}q/{\rm d}t)_{\rm AV}$ curve 
is also very abrupt (it rises by three orders of magnitude within 
$\sim$ minutes), and the peak is sharp at $\beta\gtrsim 3$ (lasting 
of the order of $\sim$ few minutes), but much softer for $\beta < 3$ 
(lasting of the order of $\sim$ an hour).
After the shock at periapsis subsides, we see the same trend as in 
the other panel: the low $\beta$ encounters result in higher post-periapsis 
shock indicators (in this case, the average $({\rm d}q/{\rm d}t)_{\rm AV}$ 
1 hour after the periapsis passage is decreasing with increasing $\beta$), 
related to long-term oscillations induced in the tidal tails.

Once the shock rebounds and travels through the star, it redistributes 
energy and angular momentum between the centre and the surface 
(the shock heating of the surface with energy from the stellar core may, 
indeed, even be responsible for a short X-ray transient, see 
\citealp{guillochon09} and \citealp{yalinewich19}). What we would 
expect to see in such a case would be a complete loss of all the 
``features'' in the energy and angular momentum histograms.
At $\beta\sim 1$, the centre of ${\rm d}M/{\rm d}\en$ is relatively 
flat (Fig.~\ref{Fig06}), corresponding to a tidal bridge whose fluid 
elements are imparted an orbital energy comparable to the one 
predicted by the frozen-in theory, while the two wings corresponding 
to the tidal lobes have a more peaked distribution); on the other hand, 
for $\beta \gtrsim 4$, where the entire star is shocked, we would 
expect the energy distribution to be more Gaussian-like.

\begin{figure*}
\centering
\includegraphics[width=\textwidth]{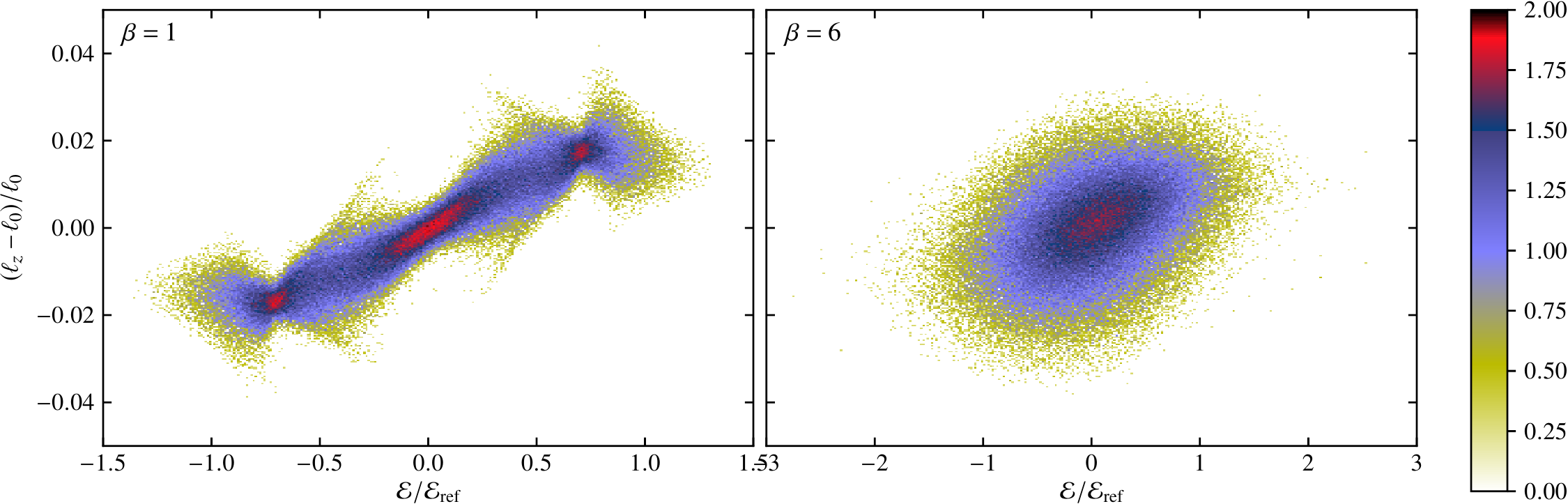}
\caption{Two-dimensional histograms of the energy $\en$ and angular 
momentum $\lz$, normalized by the reference energy spread $\en_{\rm ref}$ 
and the initial angular momentum $\ell_0$, respectively, for two Kerr 
simulations with $\aBH=0$ and impact parameters $\beta=1$ (\textit{left panel}) 
and $\beta=6$ (\textit{right panel}). The colour scheme shows the logarithm 
of the number of particles in each bin, from 0 to 2, where 300 bins have 
been used for each axis. The complete loss of features in the deep encounter 
is explained by the redistribution of energy and angular momentum by 
the shock experienced at periapsis, which leads to a Gaussian-like 
${\rm d}M/{\rm d}\en$ histogram.}
\label{Fig15}
\end{figure*}

Indeed, in Fig.~\ref{Fig15} we provide scatter  plots of the energy 
$\en$ and angular momentum $\lz$, normalized by the reference 
energy spread $\en_{\rm ref}$ and the initial angular momentum 
$\ell_0$, respectively, for two Kerr simulations with $\aBH=0$ and 
impact parameters $\beta=1$ (\textit{left panel}) and $\beta=6$ 
(\textit{right panel}). While the $\beta=1$ simulation yields a distribution 
similar to the one extracted by \citet{cheng14b}  from their relativistic 
simulations, with various features in the shape of the scatter plot, 
the $\beta=6$ simulations results in a featureless, normal distribution 
in both energy and angular momentum.

\subsection{Relativistic effects}
We analyse our results through the points of view of two opposite theoretical 
predictions. On the one hand, \citet{kesden12}, based on the model of an 
undisturbed star at periapsis (similar to the frozen-in model), argued that 
GR disruptions would be stronger for the same $\beta$, producing a larger 
spread in energies and resulting in earlier peak times and higher peak 
$\dot{M}$ rates (by up to a factor of two), more so for stars on retrograde 
orbits; our simulations do not support this conclusion. On the other hand, 
we find the argument of \citet{servin17} convincing: since the gravitational 
potential is steeper in GR (i.e., the tidal forces are stronger), the star will be 
disrupted higher in the potential well (i.e., further from the black hole), resulting 
in a smaller spread in specific energies. This creates the opposite result, of 
lower $\dot{M}$ rates, later peak times, and a broader shape for the $\dot{M}$ 
curve, in agreement with our findings and with the few previous relativistic studies 
(e.g., \citealp{cheng14b}). The two competing effects (steeper potential but earlier 
disruption) partially cancel out to yield relatively similar energy distributions and 
return rates (as seen before in Figures 9 and 11 of \citetalias{tejeda17}), making 
it more difficult to discriminate between a relativistic and a Newtonian encounter 
based solely on the mass return rate. The corrections to the shape of the debris 
fallback rate are also minor, and are strongest around $\beta\sim 4$.

On the other hand, the effect of GR on the morphology of the debris stream is 
more significant, even in weak encounters (at low $\beta$), as the differential 
periapsis shifts create much wider debris streams.
In so far as the observational prospects are concerned, we believe it much more 
likely to see the effects of GR in aspects other than the fallback rates, such as:
\begin{enumerate}
\item[(a)] the optical transient generated through recombination in the more 
expanded unbound debris stream;
\item[(b)] the bremsstrahlung radiation resulting from the breaking of the 
expanding debris stream by the ambient gas; 
\item[(c)] the $\gamma$-ray signature from the collision of the expanding 
unbound debris with molecular clouds;
\item[(d)] the X-ray signature of the shock breakout resulting from multiple 
bounces at periapsis;
\item[(e)] the longer durations of super-Eddington flow, and longer times of 
rise ($\tup$) and decay ($\tdown$); 
\item[(f)] the increased disruption rate due to the stronger disruptions at 
low $\beta$;
\item[(g)] the faster circularization, due to the self-crossing of the debris 
stream closer to the BH and at higher angles than in the Newtonian case.
\end{enumerate}

\section{Conclusions}
In this paper, we presented the results of relativistic SPH
simulations of tidal disruptions of stars by 
rotating supermassive black holes, for penetration parameters  between $0.5$ and $11$, 
and black hole spins between $-0.99$ and $0.99$.
As expected, we found that general relativity particularly 
affects deep encounters, within a few event horizon radii, as follows: 
the strong (periapsis and nodal) precession creates debris stream geometries 
impossible to obtain with the Newtonian equations (such as three-dimensional
spirals winding multiple times around the black hole, Fig.~\ref{Fig04});
part of the fluid can be launched on plunging orbits, which reduces the fallback rate 
and decreases the mass of the resulting accretion disc (by as much as 80 per cent
in the deepest encounters with retrograde spin, Fig.~\ref{Fig13});
a suite of compression and bounce episodes at periapsis in very deep relativistic
encounters (Fig.~\ref{Fig02}) may generate distinctive X-ray signatures resulting 
from the associated shock breakout;
we also found that disruptions can even occur inside the marginally bound radius, 
if the enhanced angular momentum spread launches part of the debris on 
non-plunging orbits (as is the case of all the simulations with
 $\aBH=-0.99$ and $\beta>8$).

Perhaps surprisingly, we also found relativistic effects to be important in weak disruptions, 
where the balance between self-gravity and the tidal force is very close to equilibrium. 
In this case, the otherwise minor relativistic effects can have decisive consequences 
on the qualitative outcome of the disruption.

In between, where the star is fully disrupted but relativistic effects are not extreme, 
the difference is less conspicuous and resides mostly in a gentler rise of the fallback 
rate, a later peak and a broader return rate curve, in agreement with the few 
previous relativistic simulations. However, even in the case of moderately strong encounters,
we found that the differential periapsis shift creates much thicker debris streams 
than in the Newtonian case, both in
the bound part (possibly speeding up the circularization) and in
the unbound part (speeding up the production of the recombination transient by a factor of two,
and enhancing the interaction of the ejecta with the interstellar medium).

In Appendix \ref{appendixA} we provided fit formulae for various properties of the disruption
(such as maximum compression at periapsis, shape and spread of the energy distribution) 
and potential observables (such as
 the peak fallback  rate, the times of rise, peak and decay, and the duration of the
 super-Eddington fallback phase)  as a function of the impact parameter and the black hole spin.
 
In Appendix \ref{appendixB} we discussed our strategy for binning the return times of the SPH
particles and fitting a spline to the resulting histogram, so as to generate a continuous function
$\dot{M}(t)$ based on which most quantities discussed in this paper were computed.

\section*{Acknowledgements}
The authors would like to thank John C. Miller, Elad Steinberg, and the 
anonymous referee for valuable comments on the manuscript. We would 
also like to thank Emilio Tejeda and James Guillochon for many insightful 
discussions on the topic of the paper.
S.R. has been supported by the Swedish Research Council (VR) under grant 
number 2016-03657 3, by the Swedish National Space Board under grant 
number Dnr. 107/16 and by the research environment grant ``Gravitational 
Radiation and Electromagnetic Astrophysical Transients'' (GREAT) funded 
by the Swedish Research council (VR) under Dnr. 2016-06012. Support from 
the COST Actions on neutron stars (PHAROS; CA16214) and black holes 
and gravitational waves (GWerse; CA16104) are gratefully acknowledged.
The simulations and postprocessing have in part been carried out on the 
facilities of the North-German Supercomputing Alliance (HLRN) in
Hannover, G\"ottingen and Berlin, and of the PDC Centre for High 
Performance Computing (PDC--HPC) in Stockholm.

\textit{Software}: This research has made extensive use of NASA's 
Astrophysics Data System,  \textsc{numpy} (\citealp{vanderwalt11}), 
\textsc{scipy} (\citealp{jones01}), \textsc{matplotlib} (\citealp{hunter07}) 
and \textsc{splash} (\citealp{price07b}).

\balance

\bibliographystyle{mnras}
\bibliography{paper}

\begin{thebibliography}{}
\makeatletter
\relax
\def\mn@urlcharsother{\let\do\@makeother \do\$\do\&\do\#\do\^\do\_\do\%\do\~}
\def\mn@doi{\begingroup\mn@urlcharsother \@ifnextchar [ {\mn@doi@}
  {\mn@doi@[]}}
\def\mn@doi@[#1]#2{\def\@tempa{#1}\ifx\@tempa\@empty \href
  {http://dx.doi.org/#2} {doi:#2}\else \href {http://dx.doi.org/#2} {#1}\fi
  \endgroup}
\def\mn@eprint#1#2{\mn@eprint@#1:#2::\@nil}
\def\mn@eprint@arXiv#1{\href {http://arxiv.org/abs/#1} {{\tt #1}}}
\def\mn@eprint@dblp#1{\href {http://dblp.uni-trier.de/rec/bibtex/#1.xml}
  {dblp:#1}}
\def\mn@eprint@#1:#2:#3:#4\@nil{\def\@tempa {#1}\def\@tempb {#2}\def\@tempc
  {#3}\ifx \@tempc \@empty \let \@tempc \@tempb \let \@tempb \@tempa \fi \ifx
  \@tempb \@empty \def\@tempb {arXiv}\fi \@ifundefined
  {mn@eprint@\@tempb}{\@tempb:\@tempc}{\expandafter \expandafter \csname
  mn@eprint@\@tempb\endcsname \expandafter{\@tempc}}}

\bibitem[\protect\citeauthoryear{{Anninos}, {Fragile}, {Olivier}, {Hoffman},
  {Mishra}  \& {Camarda}}{{Anninos} et~al.}{2018}]{anninos18}
{Anninos} P.,  {Fragile} P.~C.,  {Olivier} S.~S.,  {Hoffman} R.,  {Mishra} B.,
   {Camarda} K.,  2018, \mn@doi [ApJ] {10.3847/1538-4357/aadad9}, \href
  {https://ui.adsabs.harvard.edu/\#abs/2018ApJ...865....3A} {865, 3},
  \mn@eprint {arXiv} {1808.05664}

\bibitem[\protect\citeauthoryear{{Bardeen}, {Press}  \& {Teukolsky}}{{Bardeen}
  et~al.}{1972}]{bardeen72}
{Bardeen} J.~M.,  {Press} W.~H.,   {Teukolsky} S.~A.,  1972, \mn@doi [ApJ]
  {10.1086/151796}, \href {http://adsabs.harvard.edu/abs/1972ApJ...178..347B}
  {178, 347}

\bibitem[\protect\citeauthoryear{{Beck}, {Dolag}  \& {Donnert}}{{Beck}
  et~al.}{2016}]{beck16}
{Beck} A.~M.,  {Dolag} K.,   {Donnert} J.~M.~F.,  2016, \mn@doi [MNRAS]
  {10.1093/mnras/stw487}, \href
  {https://ui.adsabs.harvard.edu/\#abs/2016MNRAS.458.2080B} {458, 2080},
  \mn@eprint {arXiv} {1511.07444}

\bibitem[\protect\citeauthoryear{{Bicknell} \& {Gingold}}{{Bicknell} \&
  {Gingold}}{1983}]{bicknell83}
{Bicknell} G.~V.,  {Gingold} R.~A.,  1983, \mn@doi [ApJ] {10.1086/161410},
  \href {http://adsabs.harvard.edu/abs/1983ApJ...273..749B} {273, 749}

\bibitem[\protect\citeauthoryear{{Bogdanovi{\'c}}, {Eracleous}, {Mahadevan},
  {Sigurdsson}  \& {Laguna}}{{Bogdanovi{\'c}} et~al.}{2004}]{bogdanovic04}
{Bogdanovi{\'c}} T.,  {Eracleous} M.,  {Mahadevan} S.,  {Sigurdsson} S.,
  {Laguna} P.,  2004, \mn@doi [ApJ] {10.1086/421758}, \href
  {http://adsabs.harvard.edu/abs/2004ApJ...610..707B} {610, 707}, \mn@eprint {}
  {astro-ph/0404256}

\bibitem[\protect\citeauthoryear{{Bonnerot}, {Rossi}, {Lodato}  \&
  {Price}}{{Bonnerot} et~al.}{2016}]{bonnerot15}
{Bonnerot} C.,  {Rossi} E.~M.,  {Lodato} G.,   {Price} D.~J.,  2016, \mn@doi
  [MNRAS] {10.1093/mnras/stv2411}, \href
  {http://adsabs.harvard.edu/abs/2016MNRAS.455.2253B} {455, 2253}, \mn@eprint
  {arXiv} {1501.04635}

\bibitem[\protect\citeauthoryear{{Carter} \& {Luminet}}{{Carter} \&
  {Luminet}}{1983}]{carter83}
{Carter} B.,  {Luminet} J.-P.,  1983, A\&A, \href
  {http://adsabs.harvard.edu/abs/1983A%26A...121...97C} {121, 97}

\bibitem[\protect\citeauthoryear{{Chandrasekhar}}{{Chandrasekhar}}{1939}]{chandrasekhar39}
{Chandrasekhar} S.,  1939, {An introduction to the study of stellar structure}

\bibitem[\protect\citeauthoryear{{Chen}, {G{\'o}mez-Vargas}  \&
  {Guillochon}}{{Chen} et~al.}{2016}]{chen16}
{Chen} X.,  {G{\'o}mez-Vargas} G.~A.,   {Guillochon} J.,  2016, \mn@doi [MNRAS]
  {10.1093/mnras/stw437}, \href
  {https://ui.adsabs.harvard.edu/\#abs/2016MNRAS.458.3314C} {458, 3314},
  \mn@eprint {arXiv} {1512.06124}

\bibitem[\protect\citeauthoryear{{Cheng} \& {Bogdanovi{\'c}}}{{Cheng} \&
  {Bogdanovi{\'c}}}{2014}]{cheng14b}
{Cheng} R.~M.,  {Bogdanovi{\'c}} T.,  2014, \mn@doi [PRD]
  {10.1103/PhysRevD.90.064020}, \href
  {http://adsabs.harvard.edu/abs/2014PhRvD..90f4020C} {90, 064020}, \mn@eprint
  {arXiv} {1407.3266}

\bibitem[\protect\citeauthoryear{{Cheng} \& {Evans}}{{Cheng} \&
  {Evans}}{2013}]{cheng14}
{Cheng} R.~M.,  {Evans} C.~R.,  2013, \mn@doi [PRD]
  {10.1103/PhysRevD.87.104010}, \href
  {http://adsabs.harvard.edu/abs/2013PhRvD..87j4010C} {87, 104010}, \mn@eprint
  {arXiv} {1303.4129}

\bibitem[\protect\citeauthoryear{{Evans}, {Laguna}  \& {Eracleous}}{{Evans}
  et~al.}{2015}]{evans15}
{Evans} C.,  {Laguna} P.,   {Eracleous} M.,  2015, \mn@doi [ApJ]
  {10.1088/2041-8205/805/2/L19}, \href
  {https://ui.adsabs.harvard.edu/\#abs/2015ApJ...805L..19E} {805, L19},
  \mn@eprint {arXiv} {1502.05740}

\bibitem[\protect\citeauthoryear{{Frolov}, {Khokhlov}, {Novikov}  \&
  {Pethick}}{{Frolov} et~al.}{1994}]{frolov94}
{Frolov} V.~P.,  {Khokhlov} A.~M.,  {Novikov} I.~D.,   {Pethick} C.~J.,  1994,
  \mn@doi [ApJ] {10.1086/174607}, \href
  {http://adsabs.harvard.edu/abs/1994ApJ...432..680F} {432, 680}

\bibitem[\protect\citeauthoryear{{Gafton}, {Tejeda}, {Guillochon}, {Korobkin}
  \& {Rosswog}}{{Gafton} et~al.}{2015}]{gafton15}
{Gafton} E.,  {Tejeda} E.,  {Guillochon} J.,  {Korobkin} O.,   {Rosswog} S.,
  2015, \mn@doi [MNRAS] {10.1093/mnras/stv350}, \href
  {http://adsabs.harvard.edu/abs/2015MNRAS.449..771G} {449, 771}, \mn@eprint
  {arXiv} {1502.02039}

\bibitem[\protect\citeauthoryear{{Gezari}, {Chornock}, {Lawrence}, {Rest},
  {Jones}, {Berger}, {Challis}  \& {Narayan}}{{Gezari} et~al.}{2015}]{gezari15}
{Gezari} S.,  {Chornock} R.,  {Lawrence} A.,  {Rest} A.,  {Jones} D.~O.,
  {Berger} E.,  {Challis} P.~M.,   {Narayan} G.,  2015, \mn@doi [ApJ]
  {10.1088/2041-8205/815/1/L5}, \href
  {https://ui.adsabs.harvard.edu/\#abs/2015ApJ...815L...5G} {815, L5},
  \mn@eprint {arXiv} {1511.06372}

\bibitem[\protect\citeauthoryear{{Golightly}, {Coughlin}  \&
  {Nixon}}{{Golightly} et~al.}{2019}]{golightly19}
{Golightly} E. C.~A.,  {Coughlin} E.~R.,   {Nixon} C.~J.,  2019, \mn@doi [ApJ]
  {10.3847/1538-4357/aafd2f}, \href
  {https://ui.adsabs.harvard.edu/\#abs/2019ApJ...872..163G} {872, 163},
  \mn@eprint {arXiv} {1901.03717}

\bibitem[\protect\citeauthoryear{{Guillochon} \& {Ram\'irez-Ruiz}}{{Guillochon}
  \& {Ram\'irez-Ruiz}}{2013}]{guillochon13}
{Guillochon} J.,  {Ram\'irez-Ruiz} E.,  2013, \mn@doi [ApJ]
  {10.1088/0004-637X/767/1/25}, \href
  {http://adsabs.harvard.edu/abs/2013ApJ...767...25G} {767, 25}, \mn@eprint
  {arXiv} {1206.2350}

\bibitem[\protect\citeauthoryear{{Guillochon} \& {Ram\'irez-Ruiz}}{{Guillochon}
  \& {Ram\'irez-Ruiz}}{2015a}]{guillochon15b}
{Guillochon} J.,  {Ram\'irez-Ruiz} E.,  2015a, \mn@doi [ApJ]
  {10.1088/0004-637X/798/1/64}, \href
  {http://adsabs.harvard.edu/abs/2015ApJ...798...64G} {798, 64}

\bibitem[\protect\citeauthoryear{{Guillochon} \& {Ram\'irez-Ruiz}}{{Guillochon}
  \& {Ram\'irez-Ruiz}}{2015b}]{guillochon15}
{Guillochon} J.,  {Ram\'irez-Ruiz} E.,  2015b, \mn@doi [ApJ]
  {10.1088/0004-637X/809/2/166}, \href
  {http://adsabs.harvard.edu/abs/2015ApJ...809..166G} {809, 166}, \mn@eprint
  {arXiv} {1501.05306}

\bibitem[\protect\citeauthoryear{{Guillochon}, {Ram\'irez-Ruiz}, {Rosswog}  \&
  {Kasen}}{{Guillochon} et~al.}{2009}]{guillochon09}
{Guillochon} J.,  {Ram\'irez-Ruiz} E.,  {Rosswog} S.,   {Kasen} D.,  2009,
  \mn@doi [ApJ] {10.1088/0004-637X/705/1/844}, \href
  {https://ui.adsabs.harvard.edu/\#abs/2009ApJ...705..844G} {705, 844},
  \mn@eprint {arXiv} {0811.1370}

\bibitem[\protect\citeauthoryear{{Haas}, {Shcherbakov}, {Bode}  \&
  {Laguna}}{{Haas} et~al.}{2012}]{haas12}
{Haas} R.,  {Shcherbakov} R.~V.,  {Bode} T.,   {Laguna} P.,  2012, \mn@doi
  [ApJ] {10.1088/0004-637X/749/2/117}, \href
  {http://adsabs.harvard.edu/abs/2012ApJ...749..117H} {749, 117}, \mn@eprint
  {arXiv} {1201.4389}

\bibitem[\protect\citeauthoryear{{Hayasaki}, {Stone}  \& {Loeb}}{{Hayasaki}
  et~al.}{2013}]{hayasaki13}
{Hayasaki} K.,  {Stone} N.,   {Loeb} A.,  2013, \mn@doi [MNRAS]
  {10.1093/mnras/stt871}, \href
  {http://adsabs.harvard.edu/abs/2013MNRAS.434..909H} {434, 909}, \mn@eprint
  {arXiv} {1210.1333}

\bibitem[\protect\citeauthoryear{{Hayasaki}, {Stone}  \& {Loeb}}{{Hayasaki}
  et~al.}{2016}]{hayasaki15}
{Hayasaki} K.,  {Stone} N.,   {Loeb} A.,  2016, \mn@doi [MNRAS]
  {10.1093/mnras/stw1387}, \href
  {http://adsabs.harvard.edu/abs/2016MNRAS.461.3760H} {461, 3760}, \mn@eprint
  {arXiv} {1501.05207}

\bibitem[\protect\citeauthoryear{{Hunter}}{{Hunter}}{2007}]{hunter07}
{Hunter} J.~D.,  2007, \mn@doi [Comput. Sci. Eng.] {10.1109/MCSE.2007.55},
  \href {http://adsabs.harvard.edu/abs/2007CSE.....9...90H} {9, 90}

\bibitem[\protect\citeauthoryear{Jones, Oliphant, Peterson  et~al.}{Jones
  et~al.}{2001}]{jones01}
Jones E.,  Oliphant T.,  Peterson P.,   et~al., 2001, {SciPy}: Open source
  scientific tools for {Python}, \url {http://www.scipy.org/}

\bibitem[\protect\citeauthoryear{{Kagaya}, {Yoshida}  \& {Tanikawa}}{{Kagaya}
  et~al.}{2019}]{kagaya19}
{Kagaya} K.,  {Yoshida} S.,   {Tanikawa} A.,  2019, preprint, \href
  {https://ui.adsabs.harvard.edu/\#abs/2019arXiv190105644K} {} \mn@eprint
  {arXiv} {1901.05644}

\bibitem[\protect\citeauthoryear{{Kasen} \& {Ram\'irez-Ruiz}}{{Kasen} \&
  {Ram\'irez-Ruiz}}{2010}]{kasen10}
{Kasen} D.,  {Ram\'irez-Ruiz} E.,  2010, \mn@doi [ApJ]
  {10.1088/0004-637X/714/1/155}, \href
  {http://adsabs.harvard.edu/abs/2010ApJ...714..155K} {714, 155}, \mn@eprint
  {arXiv} {0911.5358}

\bibitem[\protect\citeauthoryear{{Kesden}}{{Kesden}}{2012}]{kesden12}
{Kesden} M.,  2012, \mn@doi [Phys. Rev. D] {10.1103/PhysRevD.86.064026}, \href
  {http://adsabs.harvard.edu/abs/2012PhRvD..86f4026K} {86, 064026}, \mn@eprint
  {arXiv} {1207.6401}

\bibitem[\protect\citeauthoryear{{Khokhlov} \& {Melia}}{{Khokhlov} \&
  {Melia}}{1996}]{khokhlov96}
{Khokhlov} A.,  {Melia} F.,  1996, \mn@doi [ApJL] {10.1086/309895}, \href
  {http://adsabs.harvard.edu/abs/1996ApJ...457L..61K} {457, L61}

\bibitem[\protect\citeauthoryear{{Kobayashi}, {Laguna}, {Phinney}  \&
  {M{\'e}sz{\'a}ros}}{{Kobayashi} et~al.}{2004}]{kobayashi04}
{Kobayashi} S.,  {Laguna} P.,  {Phinney} E.~S.,   {M{\'e}sz{\'a}ros} P.,  2004,
  \mn@doi [ApJ] {10.1086/424684}, \href
  {http://adsabs.harvard.edu/abs/2004ApJ...615..855K} {615, 855}, \mn@eprint {}
  {arXiv:astro-ph/0404173}

\bibitem[\protect\citeauthoryear{{Komossa}}{{Komossa}}{2015}]{komossa15}
{Komossa} S.,  2015, \mn@doi [J. High Energ. Astrophys.]
  {10.1016/j.jheap.2015.04.006}, \href
  {https://ui.adsabs.harvard.edu/\#abs/2015JHEAp...7..148K} {7, 148},
  \mn@eprint {arXiv} {1505.01093}

\bibitem[\protect\citeauthoryear{{Laguna}, {Miller}  \& {Zurek}}{{Laguna}
  et~al.}{1993a}]{laguna93a}
{Laguna} P.,  {Miller} W.~A.,   {Zurek} W.~H.,  1993a, \mn@doi [ApJ]
  {10.1086/172321}, \href {http://adsabs.harvard.edu/abs/1993ApJ...404..678L}
  {404, 678}

\bibitem[\protect\citeauthoryear{{Laguna}, {Miller}, {Zurek}  \&
  {Davies}}{{Laguna} et~al.}{1993b}]{laguna93}
{Laguna} P.,  {Miller} W.~A.,  {Zurek} W.~H.,   {Davies} M.~B.,  1993b, \mn@doi
  [ApJL] {10.1086/186885}, \href
  {http://adsabs.harvard.edu/abs/1993ApJ...410L..83L} {410, L83}

\bibitem[\protect\citeauthoryear{{Landau} \& {Lifshitz}}{{Landau} \&
  {Lifshitz}}{1971}]{landau71}
{Landau} L.~D.,  {Lifshitz} E.~M.,  1971, {The classical theory of fields}

\bibitem[\protect\citeauthoryear{{Leloudas} et~al.,}{{Leloudas}
  et~al.}{2016}]{leloudas16}
{Leloudas} G.,  et~al., 2016, \mn@doi [Nature Astronomy]
  {10.1038/s41550-016-0002}, \href
  {http://adsabs.harvard.edu/abs/2016NatAs...1E...2L} {1, 0002}, \mn@eprint
  {arXiv} {1609.02927}

\bibitem[\protect\citeauthoryear{{Lodato}, {King}  \& {Pringle}}{{Lodato}
  et~al.}{2009}]{lodato09}
{Lodato} G.,  {King} A.~R.,   {Pringle} J.~E.,  2009, \mn@doi [MNRAS]
  {10.1111/j.1365-2966.2008.14049.x}, \href
  {http://adsabs.harvard.edu/abs/2009MNRAS.392..332L} {392, 332}, \mn@eprint
  {arXiv} {0810.1288}

\bibitem[\protect\citeauthoryear{{Lodato}, {Franchini}, {Bonnerot}  \&
  {Rossi}}{{Lodato} et~al.}{2015}]{lodato15}
{Lodato} G.,  {Franchini} A.,  {Bonnerot} C.,   {Rossi} E.~M.,  2015, \mn@doi
  [J. High Energ. Astrophys.] {10.1016/j.jheap.2015.04.003}, \href
  {https://ui.adsabs.harvard.edu/\#abs/2015JHEAp...7..158L} {7, 158}

\bibitem[\protect\citeauthoryear{{Luminet} \& {Marck}}{{Luminet} \&
  {Marck}}{1985}]{luminet85}
{Luminet} J.-P.,  {Marck} J.-A.,  1985, \mn@doi [MNRAS]
  {10.1093/mnras/212.1.57}, \href
  {http://adsabs.harvard.edu/abs/1985MNRAS.212...57L} {212, 57}

\bibitem[\protect\citeauthoryear{{Luminet} \& {Pichon}}{{Luminet} \&
  {Pichon}}{1989}]{luminet89}
{Luminet} J.-P.,  {Pichon} B.,  1989, A\&A, \href
  {http://adsabs.harvard.edu/abs/1989A%26A...209...85L} {209, 103}

\bibitem[\protect\citeauthoryear{{Mockler}, {Guillochon}  \&
  {Ram\'irez-Ruiz}}{{Mockler} et~al.}{2019}]{mockler19}
{Mockler} B.,  {Guillochon} J.,   {Ram\'irez-Ruiz} E.,  2019, \mn@doi [ApJ]
  {10.3847/1538-4357/ab010f}, \href
  {https://ui.adsabs.harvard.edu/\#abs/2019ApJ...872..151M} {872, 151},
  \mn@eprint {arXiv} {1801.08221}

\bibitem[\protect\citeauthoryear{{Monaghan}}{{Monaghan}}{2005}]{monaghan05}
{Monaghan} J.~J.,  2005, \mn@doi [Rep. Progr. Phys.]
  {10.1088/0034-4885/68/8/R01}, \href
  {http://adsabs.harvard.edu/abs/2005RPPh...68.1703M} {68, 1703}

\bibitem[\protect\citeauthoryear{{Phinney}}{{Phinney}}{1989}]{phinney89}
{Phinney} E.~S.,  1989, in {Morris} M.,  ed.,  IAU Symp. Vol. 136, The Center
  of the Galaxy. p.~543

\bibitem[\protect\citeauthoryear{{Price}}{{Price}}{2007}]{price07b}
{Price} D.~J.,  2007, \mn@doi [PASA] {10.1071/AS07022}, \href
  {https://ui.adsabs.harvard.edu/\#abs/2007PASA...24..159P} {24, 159},
  \mn@eprint {arXiv} {0709.0832}

\bibitem[\protect\citeauthoryear{{Rees}}{{Rees}}{1988}]{rees88}
{Rees} M.~J.,  1988, \mn@doi [Nature] {10.1038/333523a0}, \href
  {http://adsabs.harvard.edu/abs/1988Natur.333..523R} {333, 523}

\bibitem[\protect\citeauthoryear{Rosswog}{Rosswog}{2009}]{rosswog09b}
Rosswog S.,  2009, \mn@doi [New Astron. Rev.] {10.1016/j.newar.2009.08.007},
  \href {http://adsabs.harvard.edu/abs/2009NewAR..53...78R} {53, 78},
  \mn@eprint {arXiv} {0903.5075}

\bibitem[\protect\citeauthoryear{{Rosswog}}{{Rosswog}}{2015}]{rosswog15c}
{Rosswog} S.,  2015, \mn@doi [Living Rev. Comput. Astrophys.]
  {10.1007/lrca-2015-1}, \href
  {http://computastrophys.livingreviews.org/Articles/lrca-2015-1/} {1, 1},
  \mn@eprint {arXiv} {1406.4224}

\bibitem[\protect\citeauthoryear{{Rosswog}, {Ram\'irez-Ruiz}, {Hix}  \&
  {Dan}}{{Rosswog} et~al.}{2008a}]{rosswog08b}
{Rosswog} S.,  {Ram\'irez-Ruiz} E.,  {Hix} W.~R.,   {Dan} M.,  2008a, \mn@doi
  [Comput. Phys. Commun.] {10.1016/j.cpc.2008.01.031}, \href
  {http://adsabs.harvard.edu/abs/2008CoPhC.179..184R} {179, 184}, \mn@eprint
  {arXiv} {0801.1582}

\bibitem[\protect\citeauthoryear{{Rosswog}, {Ram\'irez-Ruiz}  \&
  {Hix}}{{Rosswog} et~al.}{2008b}]{rosswog08}
{Rosswog} S.,  {Ram\'irez-Ruiz} E.,   {Hix} W.~R.,  2008b, \mn@doi [ApJ]
  {10.1086/528738}, \href {http://adsabs.harvard.edu/abs/2008ApJ...679.1385R}
  {679, 1385}, \mn@eprint {arXiv} {0712.2513}

\bibitem[\protect\citeauthoryear{{Rosswog}, {Ram\'irez-Ruiz}  \&
  {Hix}}{{Rosswog} et~al.}{2009}]{rosswog09}
{Rosswog} S.,  {Ram\'irez-Ruiz} E.,   {Hix} W.~R.,  2009, \mn@doi [ApJ]
  {10.1088/0004-637X/695/1/404}, \href
  {http://adsabs.harvard.edu/abs/2009ApJ...695..404R} {695, 404}, \mn@eprint
  {arXiv} {0808.2143}

\bibitem[\protect\citeauthoryear{{Runge}}{{Runge}}{1901}]{runge01}
{Runge} C.,  1901, Z. Math. Phys., 46, 224

\bibitem[\protect\citeauthoryear{{S{\c a}dowski}, {Tejeda}, {Gafton}, {Rosswog}
   \& {Abarca}}{{S{\c a}dowski} et~al.}{2016}]{sadowski16b}
{S{\c a}dowski} A.,  {Tejeda} E.,  {Gafton} E.,  {Rosswog} S.,   {Abarca} D.,
  2016, \mn@doi [MNRAS] {10.1093/mnras/stw589}, \href
  {http://adsabs.harvard.edu/abs/2016MNRAS.458.4250S} {458, 4250}, \mn@eprint
  {arXiv} {1512.04865}

\bibitem[\protect\citeauthoryear{{Servin} \& {Kesden}}{{Servin} \&
  {Kesden}}{2017}]{servin17}
{Servin} J.,  {Kesden} M.,  2017, \mn@doi [PRD] {10.1103/PhysRevD.95.083001},
  \href {http://adsabs.harvard.edu/abs/2017PhRvD..95h3001S} {95, 083001},
  \mn@eprint {arXiv} {1611.03036}

\bibitem[\protect\citeauthoryear{{Shiokawa}, {Krolik}, {Cheng}, {Piran}  \&
  {Noble}}{{Shiokawa} et~al.}{2015}]{shiokawa15}
{Shiokawa} H.,  {Krolik} J.~H.,  {Cheng} R.~M.,  {Piran} T.,   {Noble} S.~C.,
  2015, \mn@doi [ApJ] {10.1088/0004-637X/804/2/85}, \href
  {http://adsabs.harvard.edu/abs/2015ApJ...804...85S} {804, 85}, \mn@eprint
  {arXiv} {1501.04365}

\bibitem[\protect\citeauthoryear{{Steinberg}, {Coughlin}, {Stone}  \&
  {Metzger}}{{Steinberg} et~al.}{2019}]{steinberg19}
{Steinberg} E.,  {Coughlin} E.~R.,  {Stone} N.~C.,   {Metzger} B.~D.,  2019,
  \mn@doi [MNRAS] {10.1093/mnrasl/slz048}, \href
  {http://adsabs.harvard.edu/abs/2019MNRAS.485L.146S} {485, L146}, \mn@eprint
  {arXiv} {1903.03898}

\bibitem[\protect\citeauthoryear{{Stone}, {Sari}  \& {Loeb}}{{Stone}
  et~al.}{2013}]{stone13}
{Stone} N.,  {Sari} R.,   {Loeb} A.,  2013, \mn@doi [MNRAS]
  {10.1093/mnras/stt1270}, \href
  {http://adsabs.harvard.edu/abs/2013MNRAS.435.1809S} {435, 1809}, \mn@eprint
  {arXiv} {1210.3374}

\bibitem[\protect\citeauthoryear{{Stone}, {Kesden}, {Cheng}  \& {van
  Velzen}}{{Stone} et~al.}{2019}]{stone19}
{Stone} N.~C.,  {Kesden} M.,  {Cheng} R.~M.,   {van Velzen} S.,  2019, \mn@doi
  [Gen. Relativ. Gravit.] {10.1007/s10714-019-2510-9}, \href
  {https://ui.adsabs.harvard.edu/\#abs/2019GReGr..51...30S} {51, 30},
  \mn@eprint {arXiv} {1801.10180}

\bibitem[\protect\citeauthoryear{{Tejeda}, {Gafton}, {Rosswog}  \&
  {Miller}}{{Tejeda} et~al.}{2017}]{tejeda17}
{Tejeda} E.,  {Gafton} E.,  {Rosswog} S.,   {Miller} J.~C.,  2017, \mn@doi
  [MNRAS] {10.1093/mnras/stx1089}, \href
  {http://adsabs.harvard.edu/abs/2017MNRAS.469.4483T} {469, 4483}, \mn@eprint
  {arXiv} {1701.00303}

\bibitem[\protect\citeauthoryear{{Wang} \& {Merritt}}{{Wang} \&
  {Merritt}}{2004}]{wang04}
{Wang} J.,  {Merritt} D.,  2004, \mn@doi [ApJ] {10.1086/379767}, \href
  {http://adsabs.harvard.edu/abs/2004ApJ...600..149W} {600, 149}, \mn@eprint {}
  {astro-ph/0305493}

\bibitem[\protect\citeauthoryear{{Wu}, {Coughlin}  \& {Nixon}}{{Wu}
  et~al.}{2018}]{wu18}
{Wu} S.,  {Coughlin} E.~R.,   {Nixon} C.,  2018, \mn@doi [MNRAS]
  {10.1093/mnras/sty971}, \href
  {http://adsabs.harvard.edu/abs/2018MNRAS.478.3016W} {478, 3016}, \mn@eprint
  {arXiv} {1804.06410}

\bibitem[\protect\citeauthoryear{{Yalinewich}, {Guillochon}, {Sari}  \&
  {Loeb}}{{Yalinewich} et~al.}{2019}]{yalinewich19}
{Yalinewich} A.,  {Guillochon} J.,  {Sari} R.,   {Loeb} A.,  2019, \mn@doi
  [MNRAS] {10.1093/mnras/sty2809}, \href
  {https://ui.adsabs.harvard.edu/\#abs/2019MNRAS.482.2872Y} {482, 2872},
  \mn@eprint {arXiv} {1808.10447}

\bibitem[\protect\citeauthoryear{{Zheng} \& {Filippenko}}{{Zheng} \&
  {Filippenko}}{2017}]{zheng17}
{Zheng} W.,  {Filippenko} A.~V.,  2017, \mn@doi [Apj]
  {10.3847/2041-8213/aa6442}, \href
  {https://ui.adsabs.harvard.edu/\#abs/2017ApJ...838L...4Z} {838, L4},
  \mn@eprint {arXiv} {1612.02097}

\bibitem[\protect\citeauthoryear{{van der Walt}, {Colbert}  \&
  {Varoquaux}}{{van der Walt} et~al.}{2011}]{vanderwalt11}
{van der Walt} S.,  {Colbert} S.~C.,   {Varoquaux} G.,  2011, \mn@doi [Comput.
  Sci. Eng.] {10.1109/MCSE.2011.37}, \href
  {https://ui.adsabs.harvard.edu/\#abs/2011CSE....13b..22V} {13, 22},
  \mn@eprint {arXiv} {1102.1523}

\makeatother
\end{thebibliography}

{\bsp}	

\onecolumn
\appendix
\section{Fitting parameters}\label{appendixA}
We have calculated updated fitting parameters for the four characteristic 
quantities that appear in the Appendix of \citetalias{guillochon13}: the 
peak fallback rate ($\dot{M}_{\rm peak}$), the time  from the first periapsis 
passage to peak ($t_{\rm peak}$), the fraction of mass lost by the star 
($\Delta M$), and the long-term slope of the fallback rate ($n_\infty$). 
We also provide fitting parameters for several additional quantities:
the  time of rise of the fallback rate from 10 per cent of $\dot{M}_{\rm peak}$ 
to $\dot{M}_{\rm peak}$  ($\tup$), the time of decay of the fallback rate from   
$\dot{M}_{\rm peak}$ to  10 per cent of $\dot{M}_{\rm peak}$  ($\tdown$),
the duration of super-Eddington fallback ($\Delta t_{L>L_{\rm Edd}}$),
the spread in the energy distribution $\sigma_{\en}$,
and the energy distribution ${\rm d}M/{\rm d}\en$.

The fits are given in terms of functions of the penetration factor $\beta$ 
(the first four being analogous to the $A_{5/3}$, $B_{5/3}$, $C_{5/3}$ and 
$D_{5/3}$ given by \citetalias{guillochon13}), and follow the scaling with 
the  BH size $M$, the stellar mass $m_\star$ and radius $r_\star$, the 
radiative efficiency $\epsilon$ and the canonical energy spread 
$\Delta\en_{\rm ref}$ (Eq.~\ref{eq:espread} with $k_\en=1$ and $n=0$) 
given by \citetalias{guillochon13} and \citet{stone19}:
\begin{align}
\dot{M}_{\rm peak} &=
A_{5/3} \left(\frac{M}{10^6~{\rm M}_{\sun}}\right)^{-1/2}
 \left(\frac{m_\star}{{\rm M}_{\sun}}\right)^2
  \left(\frac{r_\star}{{\rm R}_{\sun}}\right)^{-3/2}~{\rm M}_{\sun}~{\rm yr}^{-1}\\
t_{\rm peak} &=
B_{5/3} \left(\frac{M}{10^6~{\rm M}_{\sun}}\right)^{1/2}
 \left(\frac{m_\star}{{\rm M}_{\sun}}\right)^{-1}
  \left(\frac{r_\star}{{\rm R}_{\sun}}\right)^{3/2}~{\rm yr}\\
\Delta M &=
C_{5/3} m\star\\
n_\infty &=
D_{5/3}\\
\tup &=
E_{5/3} \left(\frac{M}{10^6~{\rm M}_{\sun}}\right)^{1/2}
 \left(\frac{m_\star}{{\rm M}_{\sun}}\right)^{-1}
  \left(\frac{r_\star}{{\rm R}_{\sun}}\right)^{3/2}~{\rm yr}\\
\tdown &=
F_{5/3} \left(\frac{M}{10^6~{\rm M}_{\sun}}\right)^{1/2}
 \left(\frac{m_\star}{{\rm M}_{\sun}}\right)^{-1}
  \left(\frac{r_\star}{{\rm R}_{\sun}}\right)^{3/2}~{\rm yr}\\
\Delta t_{L>L_{\rm Edd}} &=
G_{5/3} \left(\frac{\epsilon}{0.1}\right)^{3/5}
\left(\frac{M}{10^6~{\rm M}_{\sun}}\right)^{-2/5}
 \left(\frac{m_\star}{{\rm M}_{\sun}}\right)^{1/5}
  \left(\frac{r_\star}{{\rm R}_{\sun}}\right)^{3/5}~{\rm yr}\\
\sigma_{\en} &=
H_{5/3} \Delta\en_{\rm ref}\\
{\rm d}M/{\rm d}\en &= J_{5/3}~10^{-4}~c^2/m_\star.
\end{align}

For $J_{5/3}$ we fit a generalized Gaussian function,
\begin{equation}\label{J53}
J_{5/3} = \frac{\mathcal{B}}{2\mathcal{A}\Gamma(1/\mathcal{B})} \exp\left[-\left(\left|\en-\bar{\en}\right|/\mathcal{A}\right)^\mathcal{B}\right],
\end{equation}
where $\bar{\en}=0$ is the expected mean specific energy after the 
disruption, $\Gamma(x)$ is the gamma function, and the parameters 
$\mathcal{A}$ and $\mathcal{B}$ are given in Table~\ref{table:fits} as 
$J_{\mathcal{A},5/3}$ and $J_{\mathcal{B},5/3}$, respectively.
Note that $\mathcal{A}^2=2\sigma^2$,  $\mathcal{B}=2$ and $\bar{\en}=\mu$ 
reduces the generalized Gaussian in Eq.~\ref{J53} to the normal distribution 
with mean $\mu$ and variance $\sigma^2$.

We point out that while the Newtonian fits are straightforward (since the fit 
functions only depend on $\beta$), in the relativistic case the problem is a 
lot more complex: not only is the dependence on $\aBH$ not strictly monotonic, 
but even in the absence of spin relativistic effects depend on the 
$r_{\rm p}/r_{\rm g}$ ratio. For instance, two encounters with the same 
$\beta$ will have very different $C_{5/3}$ curves for $M=10^6~{\rm M}_{\sun}$ 
and $M=4\times 10^7~{\rm M}_{\sun}$ (see Section 3 and in particular 
Fig.~3 of \citealp{gafton15}, where this point is clearly visible even for 
non-rotating BHs). This means that the scalings given above hold for 
Kerr only around $M\lesssim 10^6~{\rm M}_{\sun}$, and will break down 
for much larger BH masses. In so far as the deviation from Newtonian curves 
(which scale with $r_{\rm t}/r_{\rm p}$) is due to relativistic effects (which 
scale with $r_{\rm p}/r_{\rm g}$), an alternative would be to fit the relativistic 
curves in terms of offsets from the Newtonian ones, and then give those 
offsets as functions of $r_{\rm p}/r_{\rm g}$ and of $\aBH$).

For the plots in this paper, some of the time scales above were plotted in 
days, with the conversion being made assuming a Julian year, 
$1~{\rm yr}=365.25~{\rm days}$ of 86\,400 seconds each.

As opposed to \citetalias{guillochon13}, who determined their $A_{5/3}$ 
through $D_{5/3}$ via a ratio of polynomials fit, we chose to provide a 
B-spline representation of our data. This is motivated by a number of reasons: 
a) since our data extend to $\beta \gtrsim 10$, it exhibits a more complicated 
trend, with various local extrema (in particular around $\beta\sim 1$, 
$\beta\sim 4$, and $\beta\sim 11$). A higher order polynomial fit would 
be necessary to capture all of the trends in the data, making the procedure 
vulnerable to Runge's phenomenon (\citealp{runge01}); a B-spline fit 
generally avoids this issue;
b) both the coefficients and the knots of a B-spline are closely related to 
the actual data (i.e., to the $y$ and $x$ axis, respectively), while the 
coefficients of a polynomial (particularly in a ratio of polynomials) can 
grow arbitrarily high, and bear no relation to the data themselves (in 
particular, trying to fit another variable, such as the BH spin $\aBH$, 
to the coefficients, would not generally make sense);
c) we determined that a very low precision (as low as 3 significant digits 
per coefficient) is sufficient to yield a virtually indistinguishable curve 
from the one constructed with coefficients given with machine precision 
(mostly because the coefficients are related to the $y$-values); on the 
other hand, for a polynomial fit the coefficients need to be expressed 
with high precision, otherwise the fit cannot be reproduced (\citealp{guillochon15b}). 

All B-spline curves used here are cubic (i.e., of order $k=3$) and valid on a 
specific interval $[\beta_{\rm min},\beta_{\rm max}]$. In general, a B-spline 
will have an arbitrary $n_{\rm k}$ interior knots and $2(k+1)$ border knots, 
which for simplicity are normally taken as equal to the boundary values 
($\beta_{\rm min}$ and $\beta_{\rm max}$, respectively).
The number of coefficients is $n_{\rm k}+(k+1)$. For brevity, we will not 
write down the border knots (which can be deduced from the limits of the 
domain).

\begin{table}
\caption{Knots and coefficients of B-spline fits, derived separately for the 
Newtonian case ($A_{5/3,{\rm N}}$ through $J_{\mathcal{B},5/3,{\rm N}}$) 
and for the Kerr case ($A_{5/3,{\rm K}}$ through $J_{\mathcal{B},5/3,{\rm K}}$, 
as a function of the BH spin $\aBH$).}
\label{table:fits}
\begin{center}\footnotesize
\begin{tabular}{llll}
\hline
Quantity & Range of $\beta$ & Knots & Coefficients \\ \hline
$A_{5/3,{\rm N}}$ & [0.55, 11] & [0.65, 0.75, 0.85, 0.95, & 
[0.0147, 0.0201, 0.227, 0.807, 1.19, 1.36, 1.55, 1.29, 0.854, 0.578, 0.977, 1.75, 2.35]\\
&& 1.0, 1.5, 2, 4, 7]&\\
$B_{5/3,{\rm N}}$ & [0.55, 11] & Same as $A_{5/3,{\rm N}}$ & 
[0.242, 0.206, 0.18, 0.166, 0.16, 0.155, 0.154, 0.167, 0.165, 0.164, 0.0798,0.0544, 0.0433] \\
$C_{5/3,{\rm N}}$ & [0.50, 0.95] & [0.65, 0.75, 0.85] & 
[0.000242, $-0.00888$, 0.0436, 0.356, 0.696, 0.882, 0.916]\\
$D_{5/3,{\rm N}}$ & [0.55, 11] & [0.90,  0.95, 1.0, 1.4, 4] & 
[$-2.14$, $-2.42$, $-2.0$, $-1.49$, $-1.69$, $-1.66$, $-1.58$, $-1.63$, $-1.63$]\\
$E_{5/3,{\rm N}}$ & [0.55, 11] & Same as $A_{5/3,{\rm N}}$ & 
[0.0781, 0.073, 0.0674, 0.0673, 0.0714, 0.0713, 0.0799, 0.1, 0.107, 0.12, 0.0664, 0.0442, 0.0359]\\
$F_{5/3,{\rm N}}$ & [0.55, 11] & Same as $A_{5/3,{\rm N}}$ & 
[0.431, 0.401, 0.337, 0.445, 0.575, 0.61, 0.541, 0.58, 0.895, 1.49, 0.691, 0.46, 0.364]\\
$G_{5/3,{\rm N}}$ & [0.55, 11] & Same as $A_{5/3,{\rm N}}$ & 
[0.234, 0.234, 0.594, 0.815, 1.85, 2.38, 1.93, 1.96, 2.15, 2.84, 2.12, 1.76, 1.6]\\
$H_{5/3,{\rm N}}$ &  [0.55, 11] & Same as $A_{5/3,{\rm N}}$ & 
[0.497, 0.516, 0.588, 0.584, 0.551, 0.525, 0.557, 0.575, 0.522, 0.459, 0.682, 1.04, 1.25]\\
$J_{\mathcal{A},5/3,{\rm N}}$ & [1, 11] & [1.5, 2, 4, 7 ] & 
[1.95, 2.13, 2.07, 1.86, 1.11, 1.76, 2.63, 3.13]$\times 10^{-4}$\\
$J_{\mathcal{B},5/3,{\rm N}}$ & [1, 11] &[1.5, 2, 4, 7] & [3.75, 7.2, 5.22, 2.94, 1.22, 1.85, 1.63, 1.91] \\
\hline
$A_{5/3,{\rm K}}$ & [0.55, 11] & [0.65, 0.75, 0.85, 0.95, &
[$0.029-0.000402 \aBH$, $0.055-0.000316 \aBH$, $0.372-0.00438 \aBH$,
$1.04-0.0109 \aBH$\\
&&1.0, 1.5, 2, 3, 6, 9]&$1.34-0.00246 \aBH$, $1.44-0.00995 \aBH$, $1.51+0.0165 \aBH$,
$1.18-0.0132 \aBH$\\
&&&$0.836+0.0556 \aBH$, $0.548-0.0813 \aBH$, $0.673+0.108 \aBH$,
$1.58-0.278 \aBH$\\
&&&$1.88+0.565 \aBH-0.472 {\aBH}^2-0.65 {\aBH}^3$,
$1.36+4.42 \aBH-0.136 {\aBH}^2-3.75 {\aBH}^3$]\\
$B_{5/3,{\rm K}}$ & [0.55, 11] & Same as $A_{5/3,{\rm K}}$ &
[$0.226-0.0034 \aBH$, $0.2+0.00253 \aBH$, $0.178-0.000766 \aBH$,
$0.168+0.000487 \aBH$,\\
&&&$0.16-0.000565 \aBH$, $0.161-0.000215 \aBH$, $0.166-0.000977 \aBH$,
$0.176+0.00198 \aBH$,\\
&&&$0.178-0.000837 \aBH$, $0.179+0.00364 \aBH$, $0.128+0.000959 \aBH$,
$0.062+0.00872 \aBH$,\\
&&&$0.0384-0.0418 \aBH+0.00975 {\aBH}^2+0.0592 {\aBH}^3$,$0.07-0.0929 \aBH-0.0115 {\aBH}^2+0.0808 {\aBH}^3$]
\\
$C_{5/3,{\rm K}}$ & [0.50, 0.95] & [0.65, 0.75, 0.85] & [$0.00138-0.000726 \aBH$, $-0.011+0.00149 \aBH$, $0.0844-0.00348 \aBH$,\\
&&&$0.478-0.00236 \aBH$, $0.922-0.0168 \aBH$, $0.933+0.0172 \aBH$, $1+0 \aBH$]\\
$D_{5/3,{\rm K}}$ & [0.55, 11] &  [0.90,  0.95, 1.0, 1.4, 4]  &
[$-2.2 + 0.0338 \aBH$, $-2.51-0.0346 \aBH$, $-1.27+0.00813 \aBH$,
$-1.57+0.000244 \aBH$, $-1.67+$\\
&&&$0.00988 \aBH$, $-1.7-0.00489 \aBH$,
$-1.53+0.0061 \aBH$, $-1.68+0.00322 \aBH$, $-1.6-0.017 \aBH$]\\
$E_{5/3,{\rm K}}$&[0.55,11] & Same as $A_{5/3,{\rm K}}$ &
[$0.0733-0.00492 \aBH$, $0.0719+0.00329 \aBH$, $0.0672-0.00138 \aBH$,
$0.0698+0.00649 \aBH$,\\
&&&$0.0707-0.00068 \aBH$, $0.077-0.000164 \aBH$, $0.0909-0.000904 \aBH$,
$0.108+0.00186 \aBH$,\\
&&&$0.117-0.00103 \aBH$, $0.126+0.00283 \aBH$, $0.104+0.000831 \aBH$,
$0.0503+0.00677 \aBH$,\\
&&&$0.0263-0.0283 \aBH+0.013 {\aBH}^2+0.047 {\aBH}^3$,$0.0516-0.0705 \aBH-0.00674 {\aBH}^2+0.065 {\aBH}^3$]\\
$F_{5/3,{\rm K}}$&[0.55,11] & Same as $A_{5/3,{\rm K}}$ &
[$0.414+0.00304 \aBH$, $0.36-0.00437 \aBH$, $0.335+0.00363 \aBH$,
$0.492-0.00383 \aBH$,\\
&&&$0.619+0.00137 \aBH$, $0.596+0.00156 \aBH$, $0.504+0.000761 \aBH$,
$0.662-0.0057 \aBH$,\\
&&&$0.841-0.0109 \aBH$, $1.72-0.0457 \aBH$, $0.912+0.145 \aBH$,
$0.582-0.0661 \aBH$,\\
&&&$0.382+0.0969 \aBH+0.0507 {\aBH}^2+0.104 {\aBH}^3$,$0.56-0.728 \aBH-0.0328 {\aBH}^2+0.664 {\aBH}^3$]\\
$G_{5/3,{\rm N}}$&[0.55,11]& Same as $A_{5/3,{\rm K}}$ & [$0.0529-0.00482 \aBH$,
$0.234 -0.013 \aBH$, $0.572+0.0162 \aBH$, $1.06-0.0338 \aBH$, $2.54+0.00943 \aBH$,\\
&&&$2.15-0.025 \aBH$, $1.92+0.023 \aBH$, $1.99-0.0127 \aBH$,
$2.07 +0.00485 \aBH$, $2.24+0.208 \aBH$\\
&&&$2.01-0.159\aBH$, $1.52+0.686 \aBH-0.0296 {\aBH}^2 -0.247 {\aBH}^3$, 
$0.97+0.585 \aBH+0.328 {\aBH}^2 -0.128 {\aBH}^3$]\\
$H_{5/3,{\rm N}}$&[0.55,11]& Same as $A_{5/3,{\rm K}}$ &[$0.493-0.0000189 \aBH$,
$0.584-0.000538 \aBH$, $0.593-0.000516 \aBH$, $0.567+0.00131 \aBH$, $0.517+$\\
&&&$0.000487 \aBH$, $0.536-0.000206 \aBH$,
$0.558-0.00144 \aBH$, $0.553+0.00349 \aBH$, $0.521-0.00377\aBH$,\\
&&&$0.417+0.0152\aBH$, $0.574-0.0407\aBH$, $0.86+0.0224\aBH$,
$1.25-0.239\aBH$, $1.25+0.0538\aBH$]\\
$J_{\mathcal{A},5/3,{\rm K}}$ & [1, 11] &[1.5, 2, 4, 7] & 
[$2.02-0.00878 \aBH$, $2.05+0.046 \aBH$, $2.04 - 0.0726 \aBH$, $1.62+0.196 \aBH$,
$1.03-0.301 \aBH$, \\
&&&$1.98 + 0.423 \aBH$, $2.48-2.16 \aBH+0.451 {\aBH}^2 + 1.68 {\aBH}^3$,
$3.42+1.73 \aBH -0.31 {\aBH^2} - 2.21 {\aBH^3}$] $\times 10^{-4}$\\
$J_{\mathcal{B},5/3,{\rm K}}$ & [1, 11] &[1.5, 2, 4, 7] & 
[$6.44-0.107 \aBH$, $7.25+0.0387 \aBH$, $5.1-0.0428 \aBH$, 
$2.44+0.207 \aBH$, $1.32-0.185 \aBH$,\\
&&&$2.14+0.197 \aBH$, $1.38-0.978 \aBH+0.34 {\aBH}^2+0.905 {\aBH}^3$, $2.2+0.77 \aBH-0.31 {\aBH}^2-0.977 {\aBH}^3$] \\
\hline
\end{tabular}
\end{center}
\end{table}

The fits for the Newtonian and Kerr simulations are given separately in 
Table~\ref{table:fits} in terms of B-spline knots and coefficients compatible 
with popular spline functions such as \texttt{splev} from the \textsc{python} 
library \textsc{scipy}. To illustrate this, the following \textsc{python} code 
snippet generates the $\dot{M}_{\rm peak}(\beta)$ curve based on $A_{5/3}$'s 
knots and coefficients from Table~\ref{table:fits}:

\fvset{frame=single,numbers=left,numbersep=3pt,fontsize=\footnotesize}
\begin{Verbatim}
import numpy as np
from scipy.interpolate import splev
knots = [0.55, 0.55, 0.55, 0.55, 0.65, 0.75, 0.85, 0.95, 1.0, \      # Note the extra guard knots
         1.5, 2.0, 4.0, 7.0, 11.0, 11.0, 11.0, 11.0]                 # (four on each side)
coeffs = [0.0147, 0.0201, 0.227, 0.807, 1.19, 1.36, 1.55, 1.29, \
          0.854, 0.578, 0.977, 1.75, 2.35]
order = 3
xaxis = np.linspace(0.55, 11, 1000)
yaxis = splev(xaxis, [knots, coeffs, order])
\end{Verbatim}

\begin{figure*}
\includegraphics[width=\textwidth]{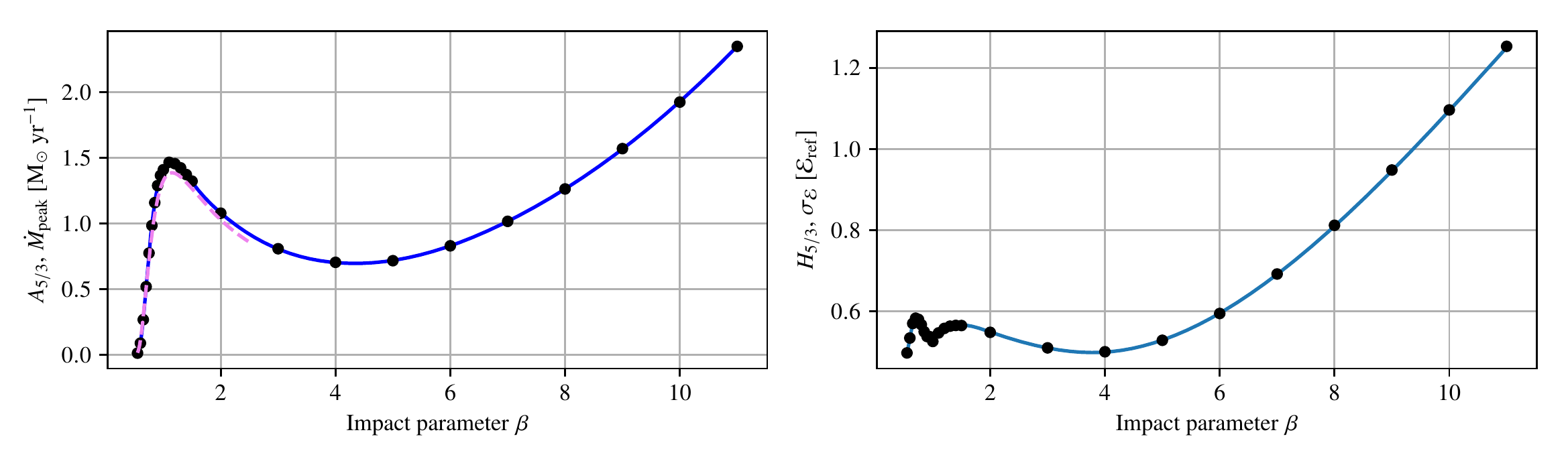}
\caption{(\textit{Left panel}) $A_{5/3}$ curve resulting from the fit given 
in Table~\ref{table:fits} (blue line), as produced by the code snippet in the 
main text, together with the $\dot{M}_{\rm peak}(\beta)$ data points 
extracted from our simulations (filled black circles) and the fits of 
\citetalias{guillochon13} (dashed purple line).
(\textit{Right panel}) $H_{5/3}$ curve resulting from the fit given in 
Table~\ref{table:fits} (blue line), together with the normalized standard 
deviation of the energy spread ($\sigma_\en/\en_{\rm ref}$)  extracted 
from our simulations (filled black circles).}
\label{FigA1}
\end{figure*}

The resulting function, defined by the arrays \texttt{xaxis} and \texttt{yaxis}, 
is shown in Fig.~\ref{FigA1} (\textit{left panel}) as a solid blue line; the data 
points from our simulations are overplotted as filled black circles, while the 
fit from \citetalias{guillochon13} is shown as a dashed purple line. In 
Fig.~\ref{FigA1} (\textit{right panel}) we repeat the same exercise, this 
time with $H_{5/3}$. In both cases, the fit to the data is excellent, and 
accomplished with as little as 3 significant digits per coefficient.

\section{Binning procedure for the fallback rate}\label{appendixB}
Once the return times $t_{\rm ret}$ to periapsis are calculated from the 
specific orbital energy (and, in the Kerr case, also from the angular momentum 
and the Carter constant, see, e.g., Appendix A of \citetalias{tejeda17}), one 
is left with a set of discrete values whose normalized histogram represents 
the mass return rate, ${\rm d}(m/m_\star)/{\rm d}t$. In order to convert the 
distribution of values $t_{{\rm ret,}i}$ (over all particles $i$) into a continuous 
function $\dot{M}(t)$ (as opposed to a discrete histogram), two steps are 
required: the binning of the data into a histogram, and the fitting of a 
sensible curve to the resulting histogram.

\subsection{Choosing the bins}\label{appendixB1}
We found that the fit of quantities such as $\dot{M}_{\rm peak}$ and 
$t_{\rm peak}$ is quite sensitive to how the data are binned and interpolated. 
In particular, both the rise part of the curve and the decay should be smooth and 
defined well enough so that one can reliably extract quantities before the peak 
(such as $\tup$) and after the peak (such as $\tdown$ and $n_\infty$). While 
one could employ statistical analysis or machine learning techniques to solve 
this problem, we have attempted to devise a very simple procedure that still 
works well at low resolution ($\sim 10^5$ particles, representing the bound 
half of the debris stream in our simulations).

\begin{figure*}
\includegraphics[width=\textwidth]{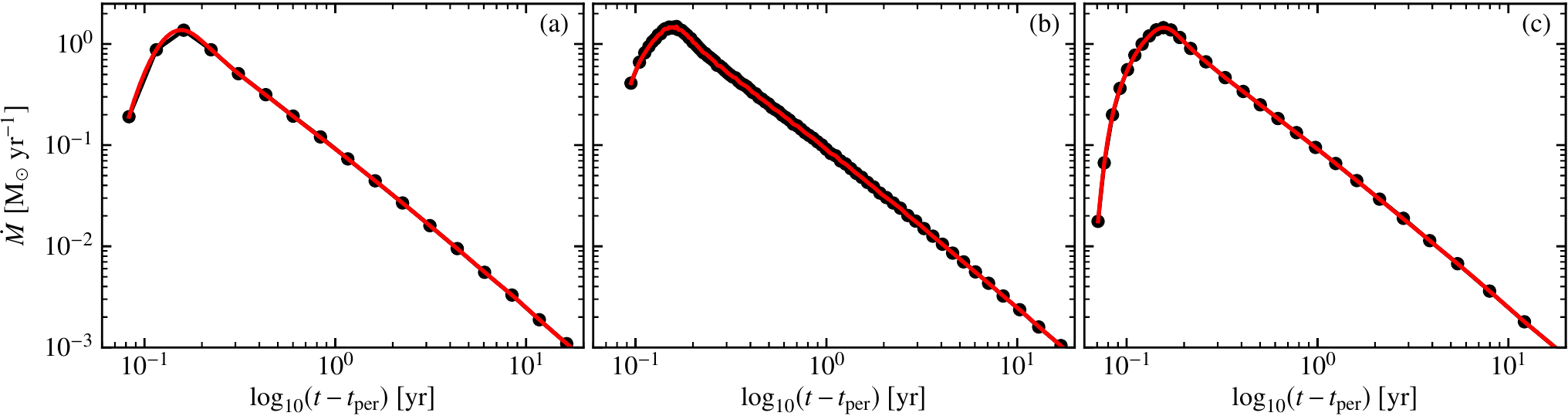}
\caption{Analysis of possible histograms of $\dot{M}(t)$ for a Newtonian 
simulation with $\beta=1$. \textit{(a)} Bins of equal width in logarithmic 
space; \textit{(b)} Bins containing an equal number of particles, here 1000; 
\textit{(c)} Bins of varying size, starting at $\sim 10$ particles at the two 
ends of the $\dot{M}$ curve and capped at $\sim 5000$ particles around 
the peak of the curve, as described in the main text. The black circles 
show the locations and values of the histogram, while the red line is a 
cubic spline fitted to the underlying histogram.}
\label{FigB1}
\end{figure*}

In order to capture the long-term evolution of the $\dot{M}$ curve, where 
both the times and the return rates can span several orders of magnitude, 
a good binning scheme should be logarithmic.
Due to finite resolution and imperfect sampling, a naive histogram (e.g., with 
equal-width logarithmic bins) will tend to be very noisy if the number of bins 
is larger than $\sim 50$, particularly in the decay part of the curve. For 
$\lesssim 50$ bins, the histogram may be quite smooth, but will fail to capture 
much of the rising part of the curve, and will also miss details of the peak 
(in particular the broken power-law typical of TDEs at $\beta\sim 1$), as 
shown in panel (a) of Fig.~\ref{FigB1}. In the case of the very low-$\beta$ 
simulations (where only a very small fraction of the star becomes bound debris), 
the effect is even worse, to the point where the resulting curve hardly resembles 
a typical TDE fallback curve (quick rise, power-law decay).

An alternative to the equal-width bins would be to use an equal number of 
particles in each bin. This would sample reasonably well the rise and decay 
part of the curve, provided the number of particles per bin is small (e.g., 1000), 
but would quickly become noisy towards the peak of the curve, where most of 
the particles are concentrated,  as shown in Fig.~\ref{FigB1}. Since these are 
competing effects, modifying the (fixed) number of particles in the bins would 
improve one of the two parts of the curve (rise+decay vs peak, i.e. low- vs 
high-resolution) at the expense of the other.

We decided, therefore, to take the best of these approaches: sample the rise 
and fall with a small number of particles per bin, and use few bins and a lot 
of particles per bin for the peak of the curve. In order to make the algorithm 
automatic and deterministic, we choose an initial number of particles 
$N_{\rm min}$ for the  first bin, at the beginning of the curve (i.e., starting 
at the particle with the shortest return time), and then successively increase 
the number of particles per bin until reaching a maximum $N_{\rm max}$. 
The same procedure is then applied starting at the last bin, at the other end 
of the curve (i.e., from the particle with the longest return time), also until 
$N_{\rm max}$ is reached. Then, the remaining particles, around $t_{\rm peak}$, 
are equally distributed into as few bins as necessary so as not to surpass 
$N_{\rm max}$ in each bin. Of course, care is taken along the way not to 
distribute more particles than available, if $N_{\rm max}$ is not reachable 
because there are very few particles, e.g., at very low $\beta$.

There are various choices of how to increase the number of particles per 
bin, of which we tested two:\\
a) multiplying the number of particles in the previous bin by a constant 
$f_{\times}$, e.g., $f_{\times}=2$, until reaching $N_{\rm max}$:
\begin{equation}
N_{\rm part,bins} = N_{\rm min}, 2 N_{\rm min}, 4 N_{\rm min}, 8 N_{\rm min}, ..., N_{\rm max}
\end{equation}
b) increasing the number of particles in the previous bin by a constant 
$f_+$, e.g., $f_+=100$, until reaching $N_{\rm max}$:
\begin{equation}
N_{\rm part,bins} = N_{\rm min}, 100 + N_{\rm min}, 200 + N_{\rm min}, 300 + N_{\rm min},  ..., N_{\rm max}
\end{equation}
The two methods give comparable results, but we have chosen b) as it 
is slightly better at sampling the low-density parts of the $\dot{M}$ curve, 
where the exponential growth of $2^n$ is too fast.

Our experiments show that, for $\sim 10^5$ particles, 
$N_{\rm min}\sim 5$, $N_{\rm max}\sim 3000$, and $f_{N+}\sim 100$ 
are good values, which minimize the noise in the histograms across the 
entire range of $\beta$. An example of such a histogram is shown in panel 
(c) of Fig.~\ref{FigB1}. It is evident even at a glance that the rise of the 
$\dot{M}$ curve is better sampled than in either of the other methods, the 
features of the peak are reproduced well, without any noise, while the 
decay part is sampled comparably to method (a), and is less noisy than  in method (b).

\subsection{Choosing a fit curve} 
\begin{figure*}
\includegraphics[width=\textwidth]{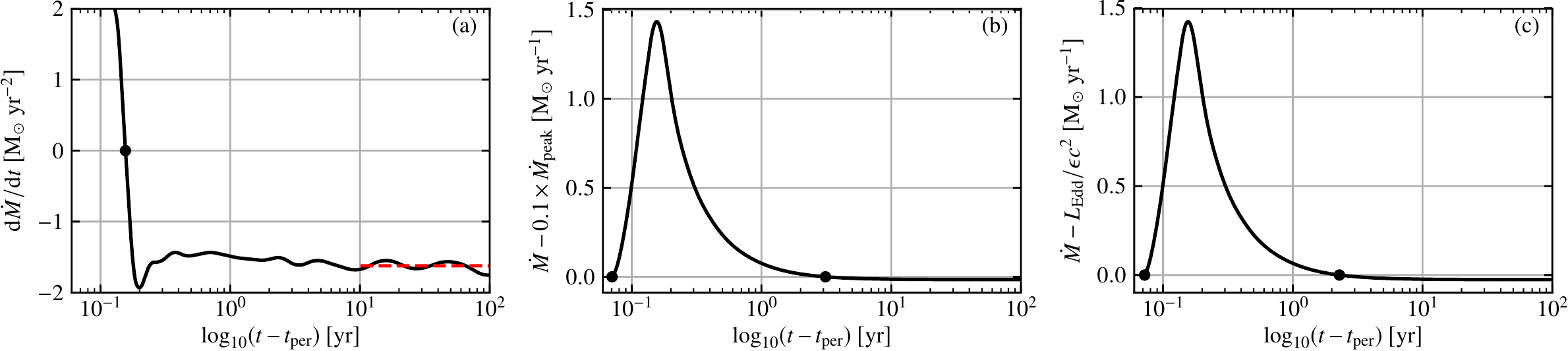}
\caption{Analysis of a cubic spline fitted to the histograms of $\dot{M}(t)$ 
for a Newtonian simulation with $\beta=1$. \textit{(a)} First derivative of 
$\dot{M}(t)$; the root of the function gives the time of the peak, $t_{\rm peak}$, 
shown here as a filled circle. The mean value of the derivative at late times 
gives the asymptotic fallback power-law decay slope $n_\infty$, shown 
here as a dashed red line; \textit{(b)} The function $\dot{M}-0.1\times \dot{M}_{\rm peak}$, 
whose roots give the quantities $\tup$ and $\tdown$, shown here as filled circles; 
\textit{(c)} The function $\dot{M}-L_{\rm Edd}/\epsilon c^2$, whose roots give 
the interval $\Delta t_{L>L_{\rm Edd}}$.}
\label{FigB2}
\end{figure*}

Once a histogram is created, we fit a cubic spline through the histogram 
points in log--log space, which gives a smooth, continuous, and -- most 
importantly -- analytically differentiable function $\dot{M}(t)$. We have 
experimented with using more complex, global functions inspired by the 
literature on supernova light curves (such as variant broken-power-law 
functions, see \citealp{zheng17}), but found them unsatisfactory in the 
case of TDEs, particularly for the rise and the decay parts (though fairly 
apt to reproduce the curve around its peak). As long as the histogram 
is carefully calculated, as detailed in Appendix~\ref{appendixB1}, a cubic 
spline generates a perfectly satisfactory differentiable function. The quantities 
of interest can then be reliably extracted using precise mathematical methods, 
typically root finding algorithms; for example, $\dot{M}_{\rm peak}$ and 
$t_{\rm peak}$ are given by the root of the first derivative of $\dot{M}(t)$, 
as illustrated in panel (a) of  Fig.~\ref{FigB2}; the quantities $\tup$ and 
$\tdown$ are given by the roots of the function 
$\dot{M}-0.1\times \dot{M}_{\rm peak}$, as illustrated in panel (b) of 
Fig.~\ref{FigB2}; $\Delta t_{L>L_{\rm Edd}}$ is given by the distance 
between the roots of the function $\dot{M}-L_{\rm Edd}/\epsilon c^2$, 
as illustrated in panel (c) of Fig.~\ref{FigB2}.

Also, the asymptotic slope $n_\infty$ is calculated as the mean derivative 
of all the splines after $t=10$ yr, as shown in panel (a) of Fig.~\ref{FigB2}. 
The cutoff is chosen somewhat arbitrarily, to ensure that: a) it is a very 
long time after the disruption, much longer than the typical time scale of 
any of the physical processes involved, and b) it is larger than both 
$t_{L=L_{\rm Edd}}$ and $\tdown$ for all of our simulations. The reason 
for taking the mean instead of the last value is to minimize the noise at 
the end of the fallback curve, where the resolution is extremely low (of 
the order of $\sim 1$ particle yr$^{-1}$) and the cubic spline is somewhat 
prone to oscillations.

\end{document}